\documentclass[superscriptaddress,
 aip,
 jmp,%
 amsmath,amssymb,
reprint,%
]{revtex4-1}

\usepackage{multirow}
\usepackage{adjustbox}
\usepackage{graphicx}
\usepackage{epstopdf}
\usepackage{dcolumn}
\usepackage{bm}
\usepackage{enumitem}
\usepackage{caption}
\usepackage{subcaption}
\usepackage{mathtools}
\usepackage{float}
\usepackage{systeme}
\usepackage{hyperref}
\usepackage{natbib}
\usepackage{xcolor}
\usepackage{soul}

\definecolor{blue}{rgb}{0,0,0}

\begin{document}

\title{Variability of echo state network prediction horizon for partially observed dynamical systems}

\author{Ajit Mahata}
\email{ajitnonlinear@gmail.com}
\affiliation{Department of Data Science, Indian Institute of Science Education and Research, IISER Pune, India 411008}

\author{Reetish Padhi}
\email{reetish.padhi@students.iiserpune.ac.in}
\affiliation{Department of Data Science, Indian Institute of Science Education and Research, IISER Pune, India 411008}
	
\author{Amit Apte}
\email{apte@iiserpune.ac.in}
\affiliation{Department of Data Science, Indian Institute of Science Education and Research, IISER Pune, India 411008}
\affiliation{International Centre for Theoretical Sciences (ICTS-TIFR), Bengaluru, India 560089}


\begin{abstract}

Study of dynamical systems using partial state observation is an important problem due to its applicability to many real-world systems. We address the problem by \textcolor{blue}{studying} an echo state network (ESN) framework with partial state input with partial or full state output. Application to the Lorenz system and Chua’s oscillator (both numerically simulated and experimental systems) demonstrate the effectiveness of our method. We show that the ESN, as an autonomous dynamical system, is capable of making short-term predictions up to a few Lyapunov times. However, the prediction horizon has high variability depending on the initial condition - an aspect that we explore in detail using the distribution of the prediction horizon. Further, using a variety of statistical metrics to compare the long-term dynamics of the ESN predictions with numerically simulated or experimental dynamics and observed similar results, we show that the ESN can effectively learn the system's dynamics even when trained with noisy numerical or experimental datasets. Thus, we demonstrate the potential of ESNs to serve as cheap surrogate models for simulating the dynamics of systems where complete observations are unavailable.

\end{abstract}


\maketitle


\section{\label{sec:intro}Introduction}

Prediction of chaotic dynamical systems is a challenging task due to exponential divergence of uncertainties in initial conditions. A variety of techniques have been developed to address this problem.~\cite{abarbanel1996analysis, kantz2004nonlinear, bradley2015nonlinear} Machine learning and neural networks have become a popular choice for forecasting such systems due to the universal approximation properties as well as computational efficiency.~\cite{schafer2007recurrent, hornik1989multilayer} 
We first review these recent approaches to the study of dynamical systems before discussing the details and novelty of the present study which aims at addressing the problem of prediction and estimation of statistical properties of dynamical systems using partial observations with the use of echo state networks. 

There have been many recent attempts at studying chaotic dynamical systems using artificial neural networks (ANN),~\cite{scher2019generalization, watson2019applying} recurrent neural networks (RNN)~\cite{728168, 1356236} and long short term memory networks (LSTM).~\cite{sangiorgio2020robustness, chattopadhyay2020data} An exhaustive study of many of these and other machine learning approaches is provided in Ref.~\onlinecite{gilpin2021chaos}.

Reservoir computing (RC)~\cite{jaeger2004harnessing, maass2002real, vogel2002computational} is one of the recent techniques that aims to address some of the problems related to requirements of large data or computational efforts and vanishing gradients~\cite{bengio2013advances} and has become an increasingly popular alternative. RC has been successfully used in various prediction tasks, namely, automatic speech recognition, ~\cite{shrivastava2021echo, skowronski2007automatic} financial time series prediction, ~\cite{lin2009short, kim2020time} natural language processing,~\cite{tong2007learning} and image identification.~\cite{schaetti2016echo} Over the years several modifications and improvements to the basic RC framework have also been proposed.~\cite{mcdermott2019deep, mcdermott2017ensemble, lokse2017training, shi2007support, lim2020predicting, lukovsevivcius2009reservoir}

In the context of dynamical systems, RC has been used to replicate attractors of chaotic systems,~\cite{zimmermann2018observing, lu2018attractor, gauthier2018reservoir, schrauwen2007overview} as well as for the problem of next-step prediction and to study the long term behaviour using correlation dimension, maximal Lyapunov exponent etc.~\cite{pathak2017using, haluszczynski2019good} using full-state of the system. A natural variation is the use of partial state of a dynamical system to assess the performance of RC, vis-a-vis other techniques such as ANN or RNN, in capturing the system dynamics.~\cite{vlachas2020backpropagation, chattopadhyay2020data, hart2020embedding} \textcolor{blue}{These studies mentioned above use RC as a tool for prediction and for studying the behavior of the dynamics and the attractor construction.~\cite{zimmermann2018observing, lu2018attractor, gauthier2018reservoir, schrauwen2007overview,pathak2017using, haluszczynski2019good} Another variation called FORCE learning technique considers the reservoir as a nonlinear dynamical system and uses the internal dynamics of the network to generate complex spatiotemporal target signal.~\cite{sussillo2009generating,maslennikov2019collective,maslennikov2023internal,masoliver2022embedded,klos2020dynamical,smith2022learning} Comparison of these various methodologies, including those discussed in the present paper, is an interesting avenue for future study.}
 
Recent theoretical results~\cite{hart2020embedding, hart2021echo} prove existence of an echo state map from a dynamical system’s phase space to the reservoir space of an appropriately constructed echo state network (ESN), leading to topologically conjugate description of the dynamics.
 
Reconstruction or prediction of the full-state dynamics using partial observation based on the RC method has been studied extensively. Most studies~\cite{nakai2018machine, song2010multi, shi2007support} use time delay embedding techniques to achieve this. We note that the time delay embedding leads to a topologically equivalent description that is not in terms of the original coordinates and usually the mapping from the original coordinates to these topologically conjugate coordinates is not learned by these techniques. On the other hand, the prediction of `unobserved' variables of the Lorenz 63 system using partial state ESN has been studied as well.~\cite{lu2017reservoir} 

Our work specifically builds on the ideas of Ref.~\onlinecite{lu2017reservoir, hart2020embedding, chattopadhyay2020data} and presents a novel framework for reconstructing the dynamics of a system from partial observations. Since no time delay embedding techniques are used, we are able to obtain the dynamics of the system in the original coordinates.

The present study is motivated by the following scenario which is commonly encountered in many applications, including those in earth sciences. Suppose we are interested in an $n$-dimensional dynamical system for which we have partial and noisy observations of $m \le n$ variables, whereas we are interested in the prediction of a larger number $l \le n$ of variables with $l \ge m$. This is quite common in most systems in earth science, including the ocean and the atmosphere - for example, we may have temperature, rainfall, and other observations at a few locations but are interested in predictions of these variables at a larger number of locations. Thus the number $m$ of observed variables is smaller than the number $l$ of variables that we want to predict, both of which are smaller than the dimension $n$ of the dynamical system itself: $m \le l \le n$.

This is exactly the setup that we investigate in this paper. \textcolor{blue}{We now give an overview of our approach, whereas the detailed mathematical setup is in} Sec.~\ref{feedback}, in particular, in discussion around Eq.~\eqref{eq:dynamical_system}-\eqref{eq-u-q-partial}. Specifically, we use an ESN with an $m$-dimensional input vector and an $l$-dimensional output vector, in order to study an $n$-dimensional dynamical system, with $m \le l \le n$. The cases with $l = n$ or $l < n$ are called, respectively, the full state output or the partial state output. The cases with $m = n$ (which of course, requires $l = n$) or $m < n$ are called, respectively, the full state input (which requires full state output) or the partial state input. The main focus of this paper (see the results in Sec.~\ref{subsec:Cha_ckt}) is on the case of partial state input and full state output, though we also provide brief comparisons with the other two cases, namely full-input, full-output (see Sec.~\ref{Sec:ESN_PH}) and partial-input, partial-output (see Sec.~\ref{subsec:partial_observation}). Note that the requirement $m \le l \le n$ implies that the fourth potential combination of full-input, partial-output is not feasible. We also note that in our setup, the ESN during the prediction phase is an autonomous dynamical system and can be used for prediction of time series of arbitrary length, instead of being restricted to making one-step predictions.

To summarize, our main objectives in this study are as follows: (a) Study chaotic dynamical systems using ESN with partial state input and full state output; (b) Evaluate the performance of ESN, with training data coming either from ODE simulation (Chua's oscillator and Lorenz 63 model) or from experimental observations (only for Chua's oscillator); (c) Use a variety of metrics that capture the performance of ESN either for a specific trajectory (mean squared error and prediction horizon) or for statistical quantities (maximal Lyapunov exponent, sample entropy, rate of growth of mean square displacement, kernel density estimates of the distributions); (d) Study of the variation of the prediction horizon with varying initial conditions; (e) Compare the performance of the ESN with partial or full state input or output.

As far as we know, no previous studies have considered ESNs trained with experimental observations using partial state input to predict full state output, although there have been studies on ESNs that are trained on experimental data from Chua's oscillator.~\cite{xi2005analyzing} In addition, one of the novelties of the present work is a detailed study of the distribution of the prediction horizon (that captures predictability) for an ensemble of different initial conditions. We note that this distribution is surprisingly wide, indicating that the predictability of the system using ESN is highly dependent on the initial conditions. This is compared with the prediction horizon obtained by perturbing the initial condition.

The rest of the paper is organized as follows. In Sec.~\ref{sec:moa}, we describe the specific architecture of ESN used in this paper followed by an introduction to the different metrics used to assess the performance of the ESN in Sec.~\ref{sec:s_test}. The specific dynamical systems used in this study are described in Sec.~\ref{sec:sss}. The main results and discussion are presented in Sec.~\ref{sec:rd}. Finally, we summarize in Sec.~\ref{sec:con} the main conclusion and indicate some directions of future research.

\section{\label{sec:moa}Methods of analysis}
We first give a general introduction to the reservoir computing framework in Sec.~\ref{subsec:method}. This is followed in Sec.~\ref{feedback} by the description of the specific way in which a part of the state vector of a dynamical system is used as input and output of the ESN in order to be able to emulate only the observed state variables using the ESN. In Sec.~\ref{heuristic}, we give a heuristic explanation for understanding why such a method may work.

\subsection{\label{subsec:method} Reservoir computing}
Reservoir computing (RC) is a machine learning method that employs the dynamics of an internal system called the reservoir to transform one time-dependent signal (input signal) into another time-dependent signal (output signal). Often, the output is just a time-delayed version of the input, thus converting RC into a tool for predicting or simulating autonomous dynamical systems. In this paper, we use the simplest form of RC architecture, the echo-state network (ESN) that was first introduced by Jaeger.~\cite{jaeger2001echo, lukovsevivcius2009reservoir}

\begin{figure}[t!]
\centering
\includegraphics[angle=0,width=8.7cm]{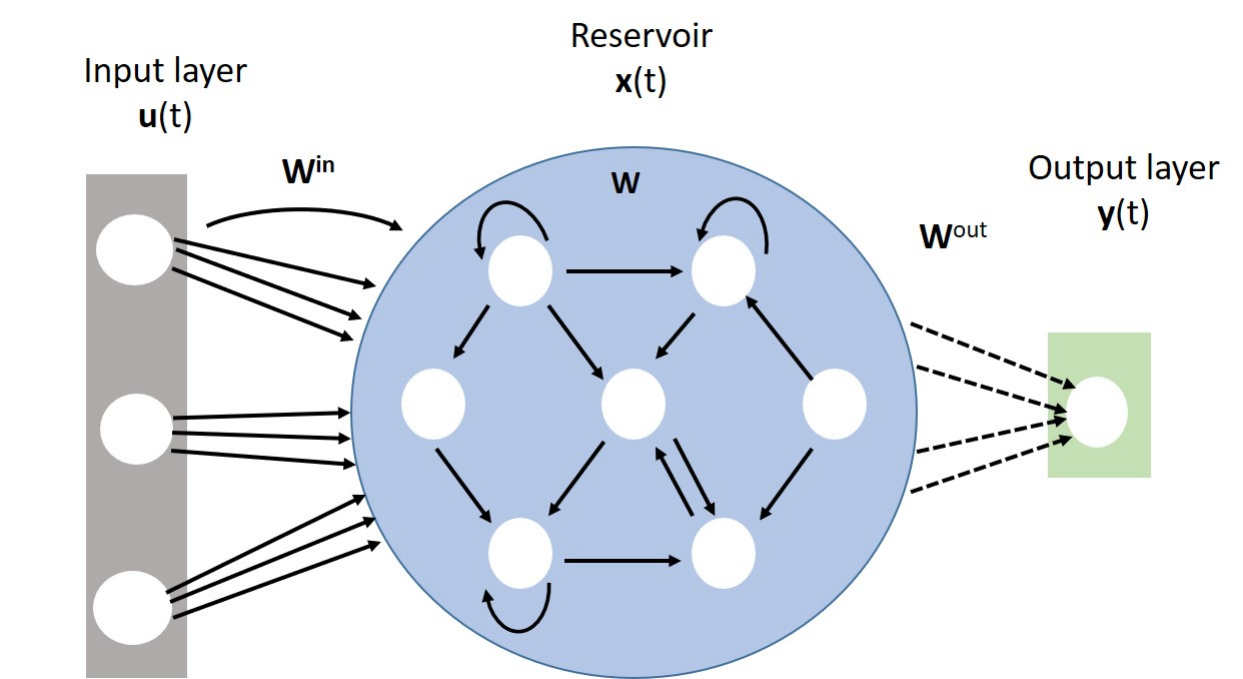}
\caption{\label{fig:ESN_atri} Reservoir computing  architecture}
\end{figure}

The general architecture of an ESN is depicted in Fig.~\ref{fig:ESN_atri}, which comprises three layers: the input layer, the reservoir, and the output layer.
The input weight matrix $\textbf{W}^{\textup{in}}$ projects the input $\mathbf{u}(t) \in \mathbb R^m$ into a higher dimensional reservoir state $\mathbf x(t) \in \mathbb R^N$. 
The reservoir consists of a dynamical system on $N$ nodes interconnected to form a randomly generated graph, described by the adjacency matrix, denoted by $\mathbf{W}$.

The reservoir state is updated at every step according to Eq.~\eqref{eq:state_eqn}.
\begin{equation}\label{eq:state_eqn}
   \mathbf x(t) = (1-\alpha) \mathbf x(t-1) + \alpha \sigma(\mathbf{W}^{\textup{in}} \mathbf u'(t)+\mathbf{W}\mathbf x(t-1)) \,,
\end{equation}
where $\mathbf u'(t)=(1,\mathbf u(t))^T\in \mathbb{R}^{1+m}$ is the combined bias and input vector at time $t$, and $\sigma: \mathbb R \to \mathbb R$ is a nonlinear function that acts on the argument component-wise. We choose $\sigma(\cdot) = \textcolor{blue}{\tanh(\cdot)}$ throughout this paper. The weight matrix $\mathbf W^{\textup{in}}$ and the adjacency matrix $\mathbf{W}$ are initialized randomly from the uniform distribution $\mathcal{U}(-0.5,~0.5)$ and remain fixed for a specific ESN. 

The output of the ESN is a linear transformation of the reservoir state and the input vector under $\textbf{W}^{\textup{out}}$. 
\begin{equation}
    \mathbf y(t) = \mathbf W^{\textrm{out}} (1,\mathbf u(t),\mathbf x(t))^\intercal \,.
\label{eq-y-def}\end{equation}
Thus given a sequence of $T$ input-output pairs $(\mathbf u(t), \mathbf y^{\textup{target}}(t))$ for $t = 1, 2, \dots, T$, and with a choice of $\mathbf x(0)$, we can think of the reservoir as a non-autonomous dynamical system that generates a sequence of states $\mathbf x(1), \mathbf x(2), \dots \mathbf x(T)$ which can then be mapped to a sequence of $T$ outputs $\mathbf y(1), \mathbf y(2), \dots \mathbf y(T)$. The ``training'' of the ESN consists of finding the output weight matrix $\mathbf W^{\textrm{out}}$ such that this ESN output sequence $\{\mathbf y(t)\}_{t=1}^T$ is ``close'' to the target output sequence $\{\mathbf y^{\textup{target}}(t)\}_{t=1}^T$.

It is common to use the $L_2$ metric to measure the distance between $\mathbf y(t)$ and $\mathbf y^{\textup{target}}(t)$. This leads us to the cost function
\begin{equation}
    J_\beta(\mathbf W^{\textrm{out}}) = \sum_{t=1}^T \left\| \mathbf y^{\textup{target}}(t) - \mathbf y(t) \right\|^2 + \beta \left\| \mathbf W^{\textrm{out}} \right\|^2
\label{eq-j-cost}\end{equation}
which is minimized with respect to $\mathbf W^{\textrm{out}}$. The second term in the cost function is the Tikhonov regularizer used to ensure convexity. Note that the first term depends on $\mathbf W^{\textrm{out}}$ through the linear dependence of $\mathbf y(t)$ on $\mathbf W^{\textrm{out}}$, as seen in Eq.~\eqref{eq-y-def}. Since $J(\cdot)$ is quadratic with respect to $\mathbf W^{\textrm{out}}$, it has a unique minimum given below. (With slight abuse of notation, we denote the minimum by the same symbol $\mathbf W^{\textrm{out}}$.)
\begin{equation}\label{eq:wout}
    \mathbf{W}^{\textup{out}} = \mathbf Y^{\textup{target}} \mathbf X^\intercal(\mathbf X \mathbf X^\intercal+\beta \mathbf I)^{-1}\,.
\end{equation}
In Eq.~\eqref{eq:wout}, $\mathbf X$ is the matrix with $j^\text{th}$ column given by $(1,\mathbf u(t),\mathbf x(t))^\intercal$ and $\mathbf Y^{\textup{target}}$ is the matrix with $\mathbf y^{\textup{target}}(t)$ as columns, for each $1 \leq t \leq T$. 

Once the output layer is ``trained,'' we fix the optimal weight matrix $\mathbf{W}^{\textup{out}}$ and use it for future predictions. For prediction, given just an input sequence $\mathbf u(t)$, the prediction time series sequence $\hat{\mathbf y}(t)$ is obtained using Eq.~\eqref{eq:esn_pred}.
\begin{equation}\label{eq:esn_pred}
    \hat{\mathbf y}(t)=\mathbf{W}^{\textup{out}}(1,\mathbf u(t),\mathbf x(t))^\intercal.
\end{equation}
Since the matrices $\mathbf{W}^{\textup{in}}$ and $\mathbf W$ are not part of the training process, it is desirable for the dynamics of the reservoir to satisfy the so-called ``echo state property" which can be stated heuristically as follows:~\cite[]{jaeger2001echo} The network state at any time $t$ is a function of the left-infinite input sequence: $\mathbf x(t) = \mathbf h(\dots, \mathbf u(t-1), \mathbf u(t))$, i.e., the semi-infinite input sequence from the past determines the present reservoir state. A sufficient condition~[Ref.\onlinecite[Proposition 3]{jaeger2001echo}] is that the largest singular value of the weight matrix $\mathbf W$ be less than one. Thus, the matrix with entries chosen randomly is simply rescaled to set the spectral radius (maximum eigenvalue) to be some desired value less than one.

The performance of the RC depends on the choice of the spectral radius $\rho$, reservoir size $N$, $\mathbf{W}^{\textup{in}}$, $\mathbf{W}$, and the parameter $\alpha$.~\cite{lukovsevivcius2012practical} The optimal values of these parameters will depend on the prediction task. We chose them through a trial and error process, so the choice is not optimal in any mathematically precise sense but rather in a heuristic sense. The problem of choosing hyperparameters in a systematic manner has been addressed in Ref.~\onlinecite{RACCA2021252}.

\subsection{\label{feedback}Using RC to reconstruct the system dynamics from partial observations} 

When the output $\mathbf y(t)$ of the ESN, or some function of it, is fed back as the input $\mathbf u(t+1)$ at the next time, the ESN acts as an autonomous dynamical system and thus can be used to predict time series of arbitrary lengths. A natural question is how well it can ``simulate'' other dynamical systems. Indeed this idea has been investigated quite extensively since the introduction of ESN in the work of Jaeger.~\cite{jaeger2001echo} Most of these studies have used ESN with both the input and output to be of the same dimension as the dynamical system being studied. The main aim of this paper is to study how well the ESN can approximate a given dynamical system in the case when we use only a part of the state space as an input and/or output to the ESN. These will be distinguished as full or partial state input and full or partial state output. We now describe the details of such a methodology in this section.

Consider an autonomous dynamical system of dimension $n$ given by Eq.~\eqref{eq:dynamical_system}.
\begin{equation}\label{eq:dynamical_system}
    \mathbf{p}(t+1)=\mathbf{g}(\mathbf p(t)) \,, \qquad \mathbf p(t) \in \mathbb R^n
\end{equation}
In this system, suppose that only $l$ out of the $n$ variables are ``observable'' with $l \le n$ and the remaining variables are unobserved. Our goal is to train an ESN that can predict the next state of these $l$ observed variables using the current state of just $m$ input variables, with $m \le l$. We assume that the observed variables correspond to the first $l$ components of $\mathbf p(t)$ by relabeling indices of $\mathbf p(t)$.

The ESN will be trained by using the observed variables of a trajectory $\mathbf p(0), \mathbf p(1), \dots, \mathbf p(T)$ of length $T+1$. In the framework introduced in the previous section, we will use the following input-output pairs for training:
\begin{align}
    &\mathbf u(t) = [p_1(t-1), p_2(t-1), \dots, p_m(t-1)]^\intercal \in \mathbb R^m \,, \nonumber \\
    &\mathbf y^{\textup{target}}(t) = [p_1(t), p_2(t), \dots, p_l(t)]^\intercal \in \mathbb R^l \,, \nonumber \\
    &\textrm{for} \quad t = 1, 2, \dots, T \,,
\label{eq-u-q-partial} \end{align}
where $p_i(t)$ is the $i$-th component of the state vector~$\mathbf p(t)$.

Once the output layer is ``trained,'' which corresponds to finding the optimal $\mathbf{W}^{\textup{out}}$ as given in Eq.~\eqref{eq:wout}, we can use the ESN for prediction of the observed variables using the following strategy. Suppose we want to predict the observed variables of a trajectory of the system~\eqref{eq:dynamical_system} with initial condition $\mathbf P = \mathbf p(T+1) \in \mathbb R^n$. Then we set the input of the ESN to be the first $m$ components of $\mathbf P$, i.e., $\mathbf u(T+1) = [P_1, P_2, \dots, P_m]^\intercal$ and we use $\mathbf x(T)$ as the state of the internal nodes. Then for every $t > T$, we can use the first $m$ components of the $l$-dimensional output $\hat{\mathbf y}(t)$ as the next input $\mathbf u(t+1)$. This is shown schematically below:
\begin{align}
    \left[ \begin{array}{c} \mathbf u(t) \\ \mathbf x(t-1) \end{array} \right]    &\xmapsto{ \mathbf{W},\mathbf{W}^{\textup{in}},\sigma\,} \left[ \begin{array}{c} \boxed{\mathbf x(t)} \end{array}\right] \,,\notag \\
    \left[ \begin{array}{c} \mathbf u(t) \\ \mathbf x(t) \end{array} \right] 
    &\xmapsto{\textup{\color{white} o} \mathbf{W}^{\textup{out}\textup{\color{white} o}}} \hat{\mathbf y}(t) \equiv \left[ \begin{array}{c} \boxed{\mathbf u(t+1)} \\ \hat{p}_{m+1}(t+1) \\ \vdots \\ \hat{p}_l(t+1) \end{array} \right] \in \mathbb R^l \,,
\label{eq:schematic} \end{align}
where the boxed quantities are then used at the next time step and the loop continues. Denoting by $\pi_m$ the projection map onto the first $m$ coordinates, we notice that ${\mathbf u}(t+1)= \pi_m \circ \hat{\mathbf y}(t)$. We also note that the whole vector $\hat{\mathbf y}(t)$ is basically the ESN prediction of the $l$ observed variables $\hat{p}_{1}(t+1), \dots, \hat{p}_{m+1}(t+1), \dots, \hat{p}_{l}(t+1)$ of the state vector $\mathbf p(t+1)$, i.e., they can be compared with $\pi_l \circ \mathbf p(t+1)$. Thus, during this prediction phase, the ESN is an autonomous dynamical system and its properties can then be compared with the original dynamical systems which generated the data used for training.

In the case when we want to use the reservoir (after training) to predict the trajectory of an initial condition different from the one that is the final training point, i.e., when $\mathbf P$ is an arbitrary point in the state space, a few time steps (denoted by $c$) will be required to initialize the reservoir. Thus, in the prediction process described above, we initialize the reservoir state to be $\mathbf x(T) = 0$ while the input is a part of the trajectory of $\mathbf P$, i.e., $\mathbf u(T + t) = \pi_m( \mathbf g^t (\mathbf P))$ for $t \le c$. The comparisons with numerically generated trajectories shown below do not include the first few steps, i.e., the comparison is done for $t > T + c$ for where $c$ is around a couple of Lyapunov times of the system.

\textcolor{blue}{The heuristic argument to understand the ESN predictions -- why the ESNs may be able to forecast the dynamical systems -- can be constructed by combining the following facts. We state this argument for the case $m = 1$.
\begin{enumerate}
    \item The `echo state' property implies that the state of the ESN at time $t$, indicated by $\mathbf x(t)$, can be written as a nonlinear function of the inputs at previous times, i.e.,
    \begin{align*}
    \mathbf x(t) = \mathbf h(\mathbf u(t),\mathbf u(t-1),\dots) \,. 
    \end{align*}
    \item By Takens' embedding theorem, there exists a function $\mathbf f$ which satisfies
    \begin{align*}
     \mathbf p(t+1) &= \mathbf f(p_1(t), p_1(t-1), p_1(t-2),\dots) \,.
    \end{align*}
    \item The output $\mathbf y(t)$ of the ESN is linearly related to its state and the input as stated in Eq.~\eqref{eq-y-def}.
\end{enumerate}
Thus, when we choose $\mathbf u(t) = p_1(t)$ as the input of the ESN and $\mathbf y(t) = \mathbf p(t+1) \in \mathbf{R}^3$ as the output, then it is clear that if the function $\mathbf h$ is such that the composition of $\mathbf h$ with this linear map is exactly the function $\mathbf f$, i.e., if $\mathbf f = \mathbf{W}^{\textup{out}} \circ \mathbf h$, then the ESN will in fact be able to predict the trajectory exactly. The training is trying to compute the $\mathbf{W}^{\textup{out}}$ to find an approximation of this equality.
A more detailed explanation is contained in Sec.~\ref{heuristic}.}

\subsection{Independance of the rows of $\mathbf{W}^\text{out}$}
We recall from Eq.~\eqref{eq:esn_pred} the expression for $\mathbf{W}^\text{out}$:
\begin{align*}      
    \mathbf{W}^\text{out} =\mathbf{Y}^\text{target} A \,,
\end{align*}
where $A = \mathbf X^\intercal(\mathbf X \mathbf X^\intercal+\beta \mathbf I)^{-1}$ which depends only on the inputs and the reservoir states. Writing $\mathbf{W}^\text{out}$ as a matrix with rows $\mathbf{w}_j^\text{out}$ for $1\leq j \leq l$, we get,
\begin{align}
    \left[\begin{array}{c}
    \mathbf{w}_1^{\text {out }} \\
    \vdots \\
    \mathbf{w}_l^{\text {out }}
    \end{array}\right] 
    =
    \left[\begin{array}{cccc}
    {y}^\text{target}_1(1) \ldots & {y}^\text{target}_1(T) \\
    \vdots & \vdots \\
    {y}^\text{target}_l(1) \ldots & {y}^\text{target}_l(T)
    \end{array}\right] A \,,
\label{eq:wout-mat} \end{align}
we notice that the expression for $\mathbf{w}^\text{out}_i$ is independent of other components $\{{y}^\text{target}_j(t)\}_{t=1}^T$ where $j\neq i$. In other words, we do not require the components $\{{y}^\text{target}_j\}_{t=1}^T$
to calculate $\mathbf{w}^\text{out}_i$ if $j\neq i$.

Consider the situation where we have two ESNs -- ESN1 and ESN2 -- both of which take a one-dimensional input $\mathbf u(t) \in \mathbb R$, but ESN1 gives output $\mathbf y_1(t) \in \mathbb{R}^3$, while ESN2 gives output $\mathbf y_2(t)\in \mathbb{R}$. Suppose we also choose $\mathbf{W}^\text{in}, \mathbf{W}, \alpha, \beta$ to be the same value for both ESNs, it then follows from Eq.~\eqref{eq:wout-mat} that $\mathbf{W}_1^\text{out}$ for ESN1 would match exactly with $\mathbf{W}^\text{out}$ of ESN2. This means that the predicted series generated by $\mathbf{W}_1^\text{out}$ of ESN1 and $\mathbf{W}^\text{out}$ should be the same. Indeed, we discuss the numerical comparison of two such ESNs in Sec.~\ref{subsec:partial_observation}.

From the above discussion, it is also clear that the prediction of the $i$-th variable $\hat p_i$ for $i > m$ is not necessary for the ESN since only the first $m$ variables are used as the next input. In other words, the case of $l > m$ and the case of $l = m$, during the prediction phase, give the same results.

\subsection{\label{heuristic}ESN as polynomial approximation}
As shown in Ref.~\onlinecite{bollt2021explaining}, an echo state network with a linear activation function can be expressed as a Vector Autoregression model (VAR). The reservoir state $\mathbf x(t)$ for a linear ESN can be written as a linear combination of $\mathbb{U}(t),\mathbb{U}(t-1),\dots,\mathbb{U}(t-k)$ where $\mathbb{U}(t)=\mathbf{W}^\textup{in}u'(t)$ as shown below, for the case $\sigma(x) = x$:
\begin{align*}
    \mathbf x(t) =& (1-\alpha)\mathbf x(t-1) + \alpha(\mathbb{U}(t) + \mathbf{W}\mathbf x(t-1)) \notag \\
     =& \mathbf M^{k}\mathbf x(t-k) + \alpha \sum \limits_{j=0}^{k-1} \mathbf M^j \mathbb{U}(t-j) \,, \notag      
\end{align*}
where $\mathbf M$ is a matrix defined such that $\mathbf M=(1-\alpha) \mathbf I + \alpha \mathbf{W}$.
As shown in Ref.~\onlinecite{jaeger2001echo}, a sufficient condition for the ESN to have the echo state property is that the spectral radius of $\mathbf M$ is less than 1. In such a case, for a sufficiently large $k$, the term $\epsilon_t = \mathbf M^{k}\mathbf x(t-k) \approx 0$ and $\mathbf x(t)$ can be written as:
\begin{equation}\label{eq:VAR_ESN}
    \mathbf{x}(t) \approx \alpha \sum \limits_{j=0}^{k-1} \mathbf M^j \mathbb{U}(t-j).
\end{equation}
Eq.~\eqref{eq:VAR_ESN_output} gives the expression for the ESN prediction $\hat{\mathbf y}(t)$ for a target variable $\mathbf y(t) \in \mathbb{R}^l$, bias-input  vector $\mathbf u'(t) \in \mathbb{R}^{1+m}$ and $\mathbf{W}^\textup{out}=(a_0,A)$, where $a_0 \in \mathbb{R}^l\times\mathbb{R}^{1+m}$ and $A \in \mathbb{R}^l\times \mathbb{R}^N$.
\begin{align}\label{eq:VAR_ESN_output}
    \hat{y}(t)&= a_0\mathbf u'(t)+A\mathbf x(t) \notag \\
    &= a_0\mathbf u'(t) + \alpha \sum_{j=0}^{k-1} A \mathbf M^j \mathbb{U}(t-j).    
\end{align}
As shown in Eq.~\eqref{eq:VAR_ESN_output}, training the ESN is equivalent to finding optimal coefficients for a linear autoregression problem. However, most functions cannot be well approximated by a linear function of past inputs. Additionally, in practice, it is seen that ESNs with non-linear activation functions such as tanh perform much better than ESNs with a linear activation function. In Eq.~\eqref{eq:Nonlinear_ESN_state}, we consider an ESN with $x-x^3/3$ (the third order power series approximation of tanh) as the activation function. Note: $A\circ B$ and $A^{\circ b}$ denote the elementwise (Hadamard) product and elementwise power of matrices respectively.
\begin{align}\label{eq:Nonlinear_ESN_state}
    \mathbf x(t) =~&(1-\alpha)\mathbf x(t-1) + \alpha(\mathbf{W}^\textup{in}\mathbf u(t) + \mathbf{W}\mathbf x(t-1))  \notag \\ & +\frac{\alpha}{3} (\mathbf{W}^\textup{in}\mathbf u(t) + \mathbf{W}\mathbf x(t-1))^{\circ 3} \notag \\ =~&\alpha\mathbb{U}(t) + \frac{\alpha}{3}\mathbb{U}(t)^{\circ 3} + \frac{\alpha}{3}(3\mathbb{U}(t)^{\circ 2}\circ \mathbb{X}(t-1) \notag \\ & + 3\mathbb{U}(t)\circ \mathbb{X}(t-1)^{\circ 2} + \mathbb{X}(t-1)^{\circ 3} ) + \mathbf M\mathbf x(t-1)\notag \\ =~&\alpha\mathbb{U}(t) + \alpha \mathbf M\mathbb{U}(t-1) + \dots \notag \\ & + \frac{\alpha}{3}\mathbb{U}(t)^{\circ 3} + \frac{\alpha}{3}\mathbf M\mathbb{U}(t-1)^{\circ 3} + \dots \notag \\ & + \alpha^2\mathbb{U}(t)^{\circ 2}\circ \mathbf{W} \mathbb{U}(t-1) + \dots \notag \\ & + \alpha^2\mathbb{U}(t)^{\circ 2}\circ \mathbf{W}\mathbb{U}(t-1)^{\circ 3} + \dots \notag \\ & + \textbf{other terms}.
\end{align}
The addition of the extra nonlinear \textcolor{blue}{${x^3}/{3}$} term results in higher order polynomial terms of $\mathbb{U}(t)$ in the expression for $\mathbf x(t)$ as shown in Eq.~\eqref{eq:Nonlinear_ESN_state}. It follows that as we take higher order polynomial terms of the tanh power series expansion, $\mathbf x(t)$ can be written as a linear combination of higher order polynomial terms of the form $\mathbb{U}(t-k)^{\circ q}$ and $\mathbb{U}(t-k)^{\circ q'}\mathbb{U}(t-k')^{\circ q''}$ for $q,q'\in \mathbb{N}$ among many others. Analogous to the linear activation case, we obtain an expression for $\hat{y}(t)$ in Eq.~\eqref{eq:Nonlinear_ESN_output}.
\begin{align}\label{eq:Nonlinear_ESN_output}
    \hat{y}(t)=& a_0\mathbf u'(t)+A\left[ \begin{array}{c} \textup{Linear combination of} \\ \mathbb{U}(t-k)^{\circ q},\mathbb{U}(t-k)^{\circ q'}\mathbb{U}(t-k')^{\circ q''} \\ \textup{ polynomial terms}\end{array} \right].
\end{align}

The rate at which the higher lag terms decay depends on the values of the spectral radius of $\mathbf M$ and choice of $\alpha$. From Eq.~\eqref{eq:Nonlinear_ESN_state} it is clear that the coefficients of $\mathbb{U}(t-k)^{\circ q}$ terms become progressively smaller as $k$ increases whenever the echo state property is satisfied. 

By the above argument, we can expect an ESN to reasonably model the partial dynamics of the system mentioned in Sec.~\ref{feedback} if given $\mathbf y(t) \in \mathbb{R}^l$ there exist $\mathbf f_1,\mathbf f_2,\dots,\mathbf f_l$ and appropriate $\delta\in \mathbb{N}$ such that Eq.~\eqref{eq:func_u} holds $\forall~ 0\leq i\leq l$. 
\begin{align}\label{eq:func_u}
    y_i(t)&=\mathbf f_i(\mathbf u(t), \mathbf u(t-1),\dots,\mathbf u(t-\delta)).
\end{align}
Thus, the output weights of a sufficiently large network ($N \gg m$) are computed such that the RHS of Eq.~\eqref{eq:Nonlinear_ESN_output} is a finite polynomial approximation, using $\mathbb{U}(t-k)$ terms, of the map $\mathbf f_i$ given in Eq.~\eqref{eq:func_u}. 

\section{\label{sec:s_test} Comparison Measures}
We use a variety of different quantities to quantify how well the ESN may capture the dynamics of a system. This section defines the various metrics used to compare the ESN predictions with the simulations of the system (using Runge-Kutta time discretization of the ODE) or with experimental data. These metrics are: prediction horizon; maximal Lyapunov exponent; asymptotic growth rate used in the 0-1 test for chaos; sample entropy; kernel density estimation.

\subsection{Prediction Horizon and Predictability}\label{ph}
Recall that the cost function defined in Eq.~\eqref{eq-j-cost} measures how far the ESN prediction is from the actual target trajectory. Considering it as a function of the two trajectories and dividing by the length of the time series and the number of dimensions, we define it as the mean squared error (recall $\mathbf y(t) \in \mathbb R^m$):
\begin{align} \label{eq:mse-def}   
    \textup{MSE}_T(\mathbf y^{\textup{target}}(t), \mathbf y(t)) &\equiv J_0(\mathbf W^{\textrm{out}})  \\
    &= \frac{1}{T \, m} \sum_{t=1}^T \left\| \mathbf y^{\textup{target}}(t) - \mathbf y(t) \right\|^2 \nonumber
\end{align}
Using the $\textup{MSE}$, the prediction horizon ${PH}_j(r)$ for a specific target trajectory indexed by $j$ and with a tolerance parameter $r$ is defined as,
\begin{equation}
    {PH}_j(r) = \inf \{T \mid \textup{MSE}_T (\mathbf y_j^{\textup{target}}(t), \mathbf y_j(t)) > r \}.
\label{eq:ph-r} \end{equation}
Simply put, the prediction horizon $PH_j(r)$ is the first time at which the mean squared error between the $j$-th target $\mathbf y_j^{\textup{target}}$ and ESN predicted series $\mathbf y_j$ exceeds a chosen threshold $r$. In this paper, the target series is either an orbit generated from an initial point by the ODE solver (RK45) or an experimentally observed time series. Since some regions on the attractor are `easier' to predict than others, it is clear that ${PH}_j(r)$ depends on the initial condition chosen to generate $\mathbf y_j^{\textup{target}}$. This means that the prediction horizon at a specific point on the attractor only describes the ESN's ability to predict the series from that point alone, making it a local quantity. To evaluate the performance of the ESN over the whole attractor, we need to consider the distribution of prediction horizon values, for a fixed value of $r$, over many initial points on the attractor. (See later Fig.~\ref{fig:PH_noise}, \ref{fig:chua_violin_PH} for examples of such distributions.) We calculate the median prediction horizon $P(r)$ of the ESN for the attractor as the median of this distribution.
\begin{equation}
    P(r) = \text{median}\{{PH}_j(r), 1 \le j \leq n\}.
\label{median-ph} \end{equation}
The reason for using the sample median and not the sample mean is that the $PH_j(r)$ distribution tends to be skewed and the median is a more reliable measure of the central tendency not affected by outliers or tails. We note that this quantity $P(r)$ is very similar to the quantity used in Ref.~\onlinecite{pathak2018hybrid} called ``valid time.'' The prediction horizon of course depends on the tolerance parameter $r$. As $r$ is arbitrary, we have calculated the prediction horizon with different values of $r$ and the variation of $PH$ is discussed in detail in Sec.~\ref{sec:rd}.

\subsection{\label{MLE}Maximal Lyapunov Exponent}
Lyapunov exponents for a dynamical system describe the rate of separation of two infinitesimally close trajectories. Heuristically, if $\delta x(0)$ is the initial separation between two nearby phase space trajectories, then their separation at time $t$ would be $\delta x(t)\approx \delta x(0) e^{\lambda t}$, where $\lambda$ is the Lyapunov exponent. The rate of separation can vary depending on the orientation of the initial separation vector. As a result, there is a spectrum of Lyapunov exponents proportional to the dimensionality of the phase space. The largest of these exponents is commonly referred to as the maximal Lyapunov exponent (MLE). A positive MLE is an indication that the system is chaotic. The details of the method we use for estimating MLE can be found in Ref.~\onlinecite{rosenstein1993practical}.  

It should be noted that an arbitrary initial separation vector will typically contain some component in the direction associated with the MLE, and the effect of the other exponents will be obliterated over time due to the exponential separation rate. The characteristic timescale of a dynamical system is measured in terms of its Lyapunov time which is equal to the inverse of the MLE. Consequently, the MLE of a dynamical system is a valuable metric for describing the long-term behaviour of a system. Hence we use the MLE of the ESN predicted and the original time series as one of the measures of the ESN performance. Further, all results (such as the prediction horizon) in the subsequent sections are reported in terms of the Lyapunov time since it allows us to compare results across different dynamical systems. 

\subsection{\label{0_1_test}0-1 Test for Chaos}
The 0-1 test~\cite{gottwald2009implementation} is a binary test used to distinguish between regular and chaotic dynamics. In order to determine if a time series is chaotic or not, it can be used as a helpful confirmation test. Unlike the calculation of the Lyapunov exponent, this method does not require the use of reconstruction methods. The details of the test are discussed in Ref.~\onlinecite{gottwald2009implementation}. We use the asymptotic growth rate $K_c$ of the mean square displacement as another measure of evaluating the performance of the ESN.

\subsection{\label{Sampen}Sample Entropy}
The sample entropy (SE) of a time series is defined as the negative natural logarithm of the conditional probability that two sequences similar for $N$ points remain identical at the next point, excluding self-matches. A lower SE value corresponds to a higher probability that two similar sequences remain similar at the next point. SE is a commonly used method to quantify the complexity and irregularity of time series data generated by dynamical systems. Therefore, it serves as a useful metric to compare the similarity of the predicted and original sequences. The details of the method of estimation of SE can be found in Ref.~\onlinecite{delgado2019approximate}.
\subsection{\label{Karnel_den}Kernel density estimation}
A kernel density estimation (KDE) plot is a useful tool for visualizing the distribution of a data set. In our study, we used KDE for the comparison between the ESN-predicted time series and the simulated or experimental time series. The details of KDE can be found in Ref.~\onlinecite{vermeesch2012visualisation}.

\section{\label{sec:sss} Systems selected for the study}
We have selected Lorenz 63 and Chua's circuit for this study. We use the standard Lorenz 63 system given by the following system of equations: 
\begin{align} 
    \dot x &= \sigma(y-x)\,, \nonumber\\ 
    \dot y &= x(\rho - z) - y\,, \nonumber\\
    \dot z &= xy-\beta z\,.
\label{eq-l63} \end{align}
with the standard choice of parameters: $\sigma=10,\rho=28,\beta=8/3$.

\begin{figure*}[t!]
\centering
\includegraphics[angle=0,width=16cm]{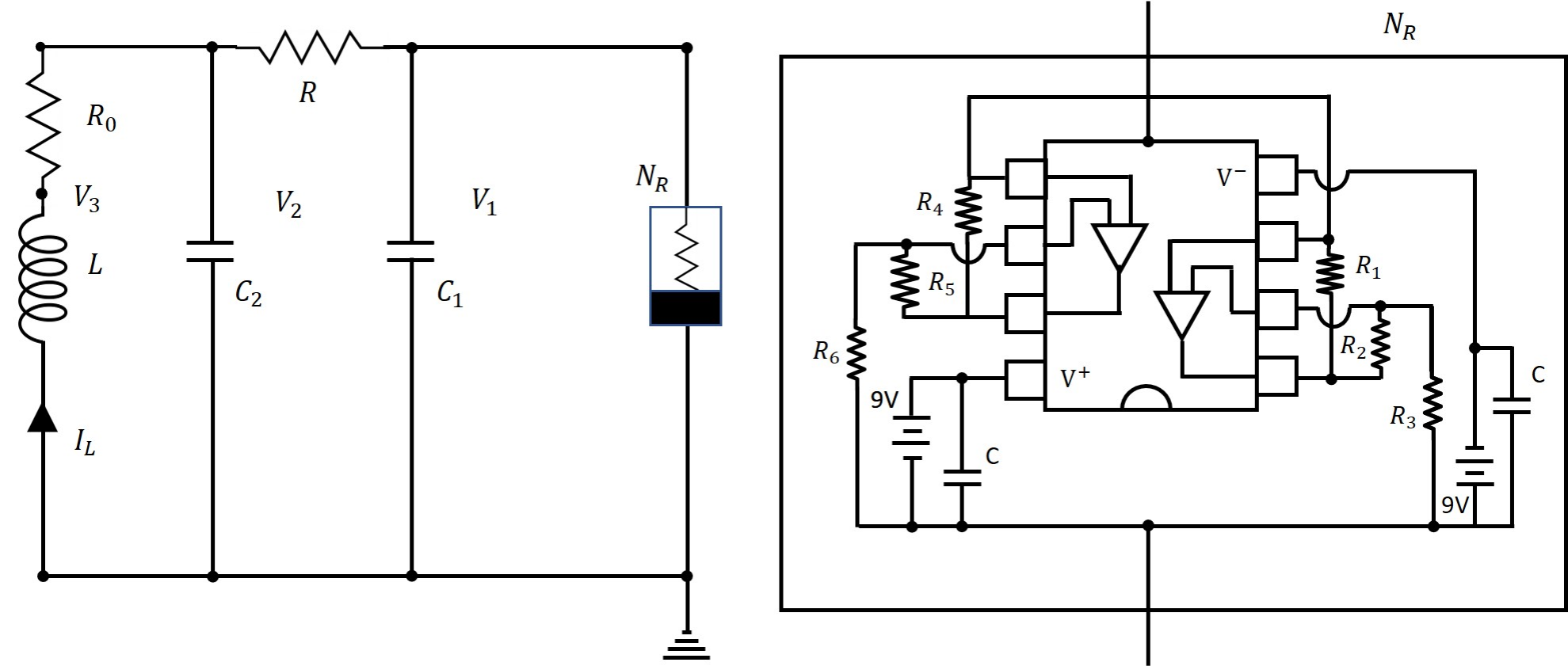}
\caption{\label{fig:Chua_experiental_setup} Schematic diagram of the Chua's circuit experimental set-up (left panel), along with the details of the nonlinear resistor $(N_R)$ (right panel).}
\end{figure*}
Fig.~\ref{fig:Chua_experiental_setup} shows Chua's oscillator circuit that we use, both in the simulation and in the experiments. The circuit contains linear capacitors and inductor, and linear and nonlinear resistors, with the nonlinear resistor constructed using OP-AMP. The dynamics of the system can be written as follows:
\begin{align*} 
    C_1 V'_1 &=  (V_2-V_1)/R-g(V_1) \,,\\ 
   C_2 V'_2 &= (V_1-V_2)/R+I_L \,,\\
    L I'_L &= -V_2-R_0I_L \,.
\end{align*}
where the prime denotes $d/d\tau$ with $\tau$ being time in seconds, $g(V_R)$ is the current through the nonlinear resistor given by
\begin{align*}
   g(V_R)&= G_bV_R+0.5(G_a-G_b)(|V_R+B_p|-|V_R-B_p|)\,.
\end{align*}
Here, $G_b$ and $G_a$ are the slopes of the outer and inner regions of the current-voltage graph. $+B_p$ and $-B_p$ are the breakpoints of the graph.

We converted this equation into non-dimensional form by considering $x=V_1/B_p$, $y=V_2/B_p$, $z=RI_L/B_p$, and $t = \tau/RC_2$. So, the non-dimensional form of the Chua's circuit is given by the following system of equations:
\begin{align} 
    \dot x &= \alpha(y-x-\phi(x))\,, \nonumber\\ 
    \dot y &= x - y + z\,, \nonumber\\
    \dot z &= -\beta y - \gamma z\,.
\label{eq-chua-nondim} \end{align}
where $\phi(x)$ is the non linear function given by
\begin{align*}
    \phi(x) &= m_1x + \frac{1}{2}(m_0-m_1)(|x+1| - |x-1|)\,.
\end{align*}
The definitions and the values of the various non-dimensional parameters that we use are as follows: $\alpha=C_2/C_1 = 10.0$, $\beta=R^2C_2/L = 9.77$, $\gamma=RR_0C_2/L = 0.58$, $m_0=RG_a = -0.735$ and $m_1=RG_b = -1.301$. \textcolor{blue}{In this study, we selected a reservoir size of 1000 for Chua's oscillator and 3000 for the Lorenz 63 model. A brief discussion on the effect of reservoir size on the prediction horizon distribution has been provided in the appendix~\ref{appendix:size_ESN}}.

\subsection {Experimental setup of Chua's circuit}\label{ssec:expt-chua}
We have constructed Chua's circuit in an experimental set-up with the schematic diagram shown on the left side in Fig.~\ref{fig:Chua_experiental_setup}. TL082 OP-AMP and linear resistors $R_0$, $R_1$, $R_2$, $R_3$, $R_5$ and $ R_6$ have been used to construct the nonlinear resistor $N_R$ acting as Chua's diode, which is shown in the right side of Fig.~\ref{fig:Chua_experiental_setup}. The values of the inductor, resistors and capacitors have been chosen in the following way: $R_0=83~\Omega$, $R_1=220~\Omega$, $R_2=220~\Omega$, $R_3=2.2~K\Omega$, $R_4=22~K\Omega$, $R_5=22~K\Omega$, $R_6=3.3~K\Omega$, $C_1=10~nF$ and $C_2=100~nF$. In the experimental setup, $R$ is varied by using a 5 $K\Omega$ potentiometer in order to get different types of dynamics - chaotic or regular. The voltages $V_1, V_2$ across the capacitors $C_1, C_2$ are measured directly. To measure the current through the inductor, we measure the voltage across the inductor $V_3$, then calculate the voltage $V_0$ across the resistance $R_0$ to be $V_0=(V_2-V_3)$, and finally, the current through the inductor is calculated as $I_L=V_0/R_0$. In this set-up, we observe that at $R=1.398~K\Omega$, the dynamics show the double scroll attractor. In this condition, we collected the $I_L$, $V_2$, and $V_1$ time series to study the dynamics using reservoir computing.

\section{\label{sec:rd}Results and discussion}
As mentioned earlier, the main focus of this paper is on evaluating the performance of ESN with partial state input (namely, one-dimensional input) and full state output (namely, three-dimensional output). These results are discussed in the main Sec.~\ref{subsec:Cha_ckt} below. We also implemented the variation of the ESN where the input is the three-dimensional full-state vector, and the comparison with the partial state input is discussed in Sec.~\ref{Sec:ESN_PH}. The other variation we discuss is the case when the output is a one-dimensional partial state, with the results in this case being discussed in Sec.~\ref{subsec:partial_observation}.

\subsection{\label{subsec:Cha_ckt} Study of Chua's oscillator and Lorenz 63 dynamics: ESN with full state output}

This section discusses all the results for the following setup: the ESN input is one-dimensional (partial state), while the output is three-dimensional (full state). The three systems studied are: Chua oscillator ODE simulations, Lorenz 63 ODE simulations, and experimental data from the Chua circuit setup described in Sec.~\ref{ssec:expt-chua}. In the notation of Eq.~\eqref{eq-l63} or~\eqref{eq-chua-nondim}, the network was trained to predict all three variables $x(t+1), y(t+1), z(t+1)$ using just $x(t)$ as input. Thus in the notation of Sec.~\ref{feedback}, $\mathbf u(t) \in \mathbb R$ is $x(t)$ of Eq.~\eqref{eq-l63} or~\eqref{eq-chua-nondim} and $\mathbf y(t) \in \mathbb R^3$ is $(x(t), y(t), z(t))$. 

We note that the ODE simulations use the dimensionless form of equations and corresponding dimensionless data for ESN training and testing while in the case of the experimental system, the ESN is trained and tested against the dimensional variables of voltage and current $(V_1, V_2, I_L)$ in the units of $V$ and $A$, instead of $(x, y, z)$.

Chua's circuit equations were solved using RK45 with $\Delta t=0.05$ to generate a dataset of 75000 datapoints ($\approx$ 400 Lyapunov times). Similarly, Lorenz 63 equations were solved using RK45 with $\Delta t=0.01$ to generate a dataset of 100000 datapoints ($\approx$ 950 Lyapunov times). For ESN trained on experimental data, we used $\Delta t = 0.057$ (non-dimensional, corresponding to $\Delta \tau = 8 \times 10^{-6} s$ and a dataset of 120000 datapoints ($\approx$ 750 Lyapunov times).

In all these cases, the ESN was trained on 80\% of the dataset, while the rest was reserved for testing. After training, a single initial condition (one component of the full state vector) along with the reservoir state was used to generate, following the method shown in Eq.~\eqref{eq:schematic}, a series of predictions, which are then compared with the test series obtained by numerical solution of the ODE.

We first present both qualitative (Sec.~\ref{subsec:chua_shrt_prediction}) and quantitative (Sec.~\ref{ssec-stat-char-ode}) evaluation of the ESN performance, followed by a detailed discussion of the variability of the prediction horizon for different trajectories (Sec.~\ref{ssec-mse-ph} and Sec.~\ref{sec:uncertainty_initcond}). We examine the effects of noisy data (section~\ref{ssec-noise}) and then the results of ESN trained with experimental data (Sec.~\ref{subsec:experiment_chua's}).

\subsubsection{\label{subsec:chua_shrt_prediction} Forecasting of Chua and Lorenz 63 dynamics}
\begin{figure}[t!]
\centering
\includegraphics[angle=0,width=8.5cm]{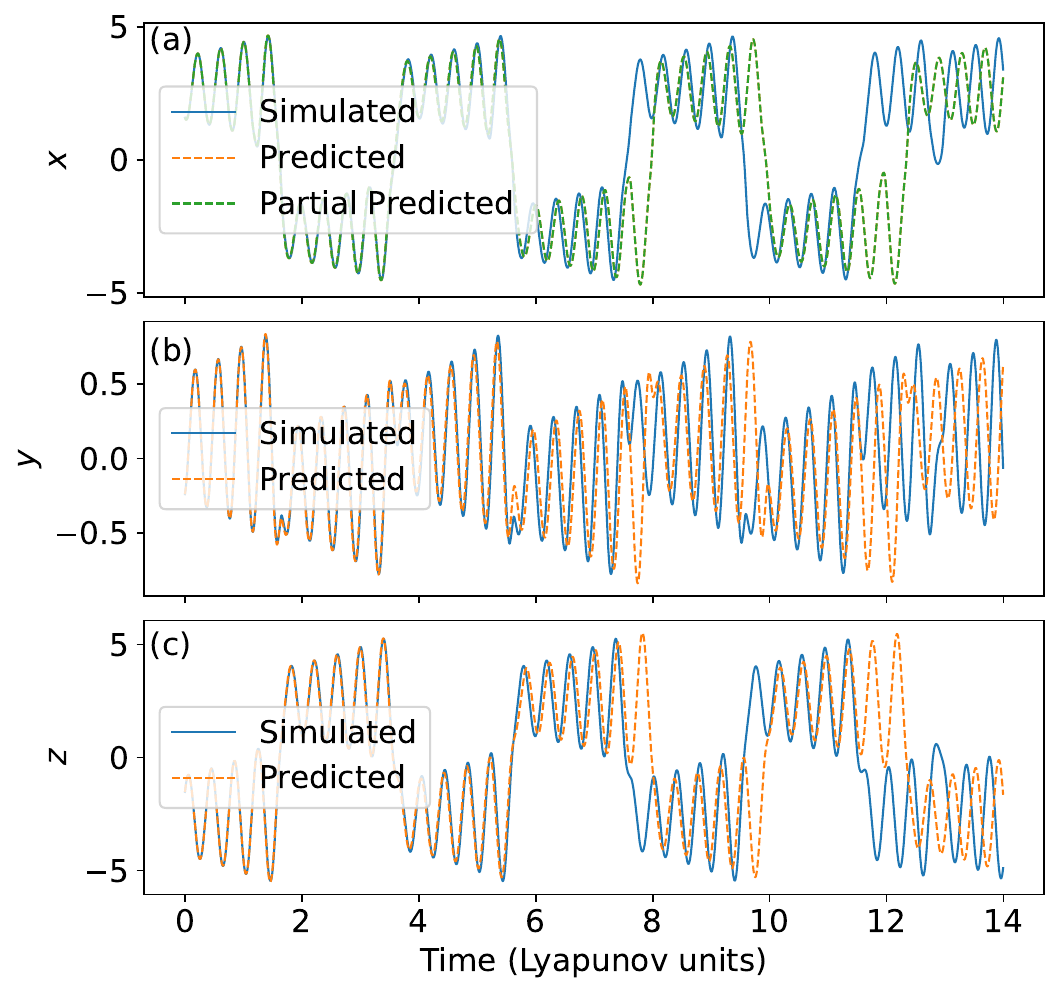}
\caption{\label{fig:chua_test_pred} Comparison of a numerically simulated (solid) trajectory with the output of the ESN with 3-dimensional full state output (thin dashed) and  1-dimensional partial state output (thick dashed) for Chua's oscillator.}
\end{figure}
Fig.~\ref{fig:chua_test_pred} illustrates a typical numerically simulated (solid line) and ESN predicted (dashed line) time series of the non-dimensional form of $x$, $y$, and $z$ variables of the Chua's oscillator. The ESN can make successful short-term predictions but eventually diverges, as expected for a chaotic system. 

The time up to which the ESN predictions are close to the original trajectory is very different for different trajectories. In other words, the prediction horizon, defined in Eq.~\eqref{eq:ph-r}, is highly dependent on the trajectory. A more extensive discussion of the full distribution for the prediction horizon is presented in Sec.~\ref{Sec:ESN_PH}. It is found that the median prediction horizon, as defined in Eq.~\eqref{median-ph}, for the ODE simulated Chua's circuit data with a threshold of $0.01$, i.e., $P(0.01)$, is 1.87 Lyapunov times. (See also Table~\ref{tab:TABLE2}.) 

\begin{figure*}[t!]
\centering
\includegraphics[angle=0,width=17.5cm]{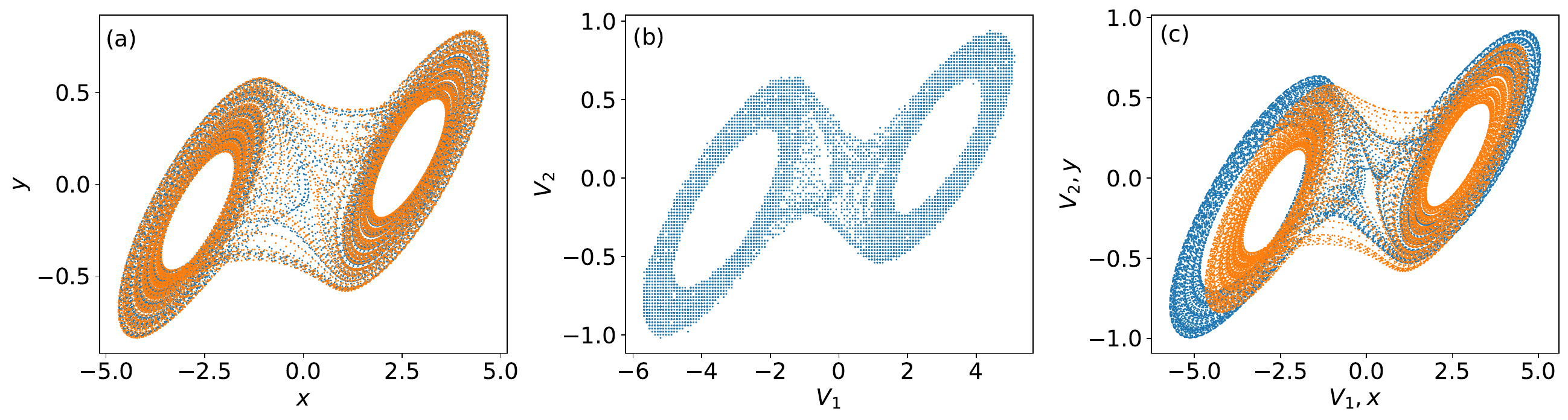}
\caption{\label{fig:chua_attr} Chua's oscillator: Blue and orange colors in (a) show, respectively, the attractors using ODE simulations and using ESN trained with ODE simulation data. (b) shows the attractor using the experimental data. Blue and orange colors in (c) show, respectively, the attractors using ESN trained with ODE simulation data and ESN trained with experimental data, showing that the experimental and ODE simulated attractors are not similar to each other.}
\end{figure*}
Fig.~\ref{fig:chua_attr}(a) blue and orange colors show the double scroll attractor of Chua's oscillator projected to the $x$-$y$ plane for the simulated and the ESN predicted series, respectively. We see that both attractors are very similar in nature. 

\begin{figure}[t!]
\centering
\includegraphics[angle=0,width=8.5cm]{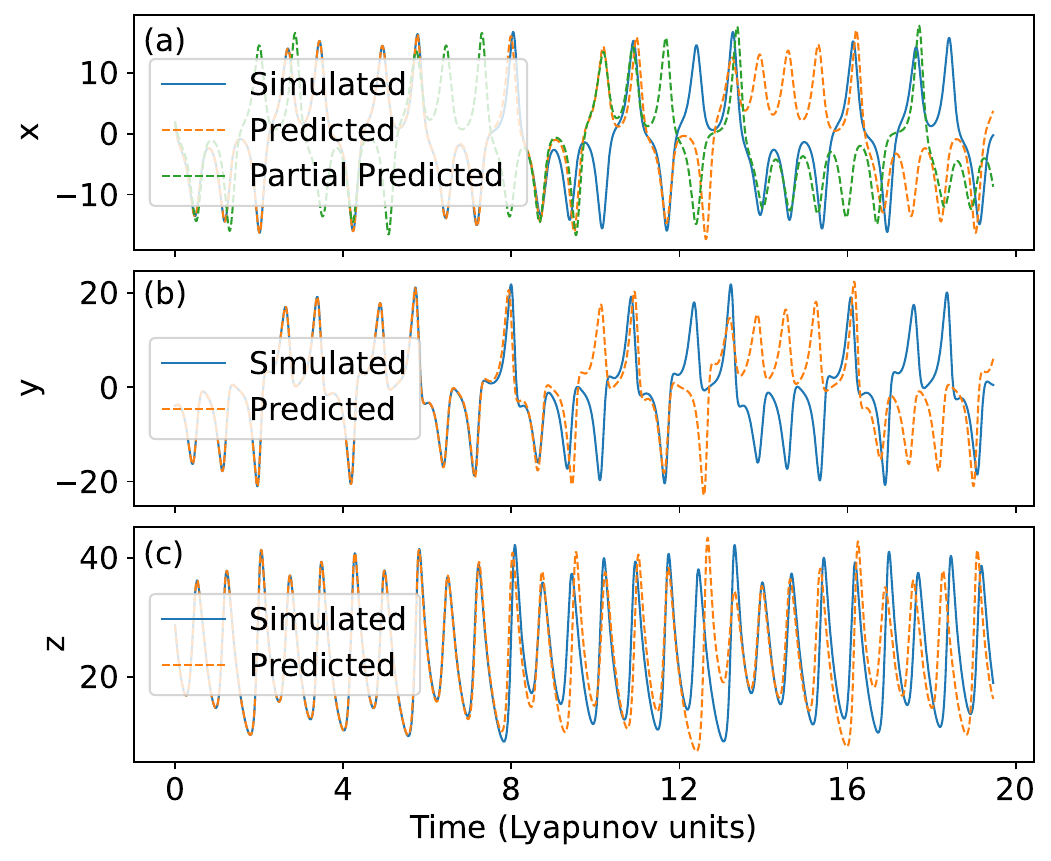}
\caption{\label{fig:Lorenz_Shrt_Pred} Same as Fig.~\ref{fig:chua_test_pred} but for Lorenz 63 model.}
\end{figure}
Fig.~\ref{fig:Lorenz_Shrt_Pred} shows the simulated (solid) and predicted (thin dashed) time series of $x$, $y$ and $z$ of the Lorenz 63 system, respectively. The median prediction horizon~$P(0.01)$ for the Lorenz system is 4.105 Lyapunov times, very similar to the Chua oscillator case. 
\begin{figure}[t!]
\centering
\includegraphics[angle=0,width=8.5cm]{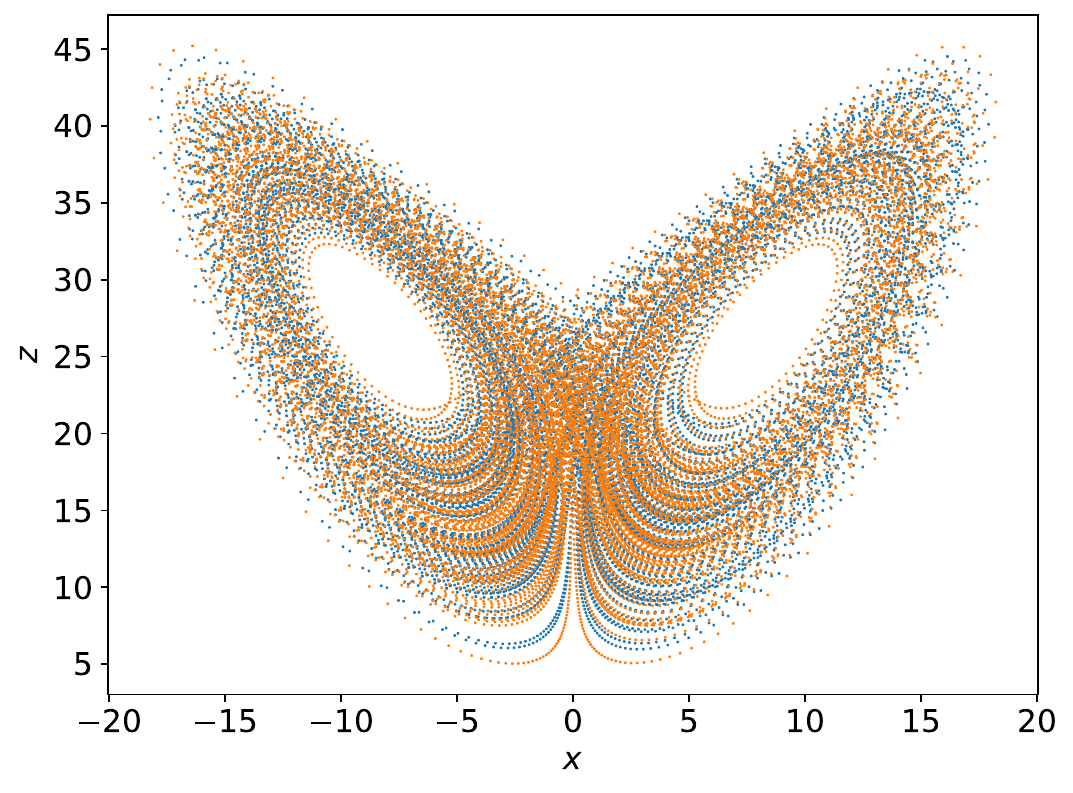}
\caption{\label{fig:Lorenz_Att} Blue and orange colors represent the simulated and ESN predicted for the Lorenz 63 model.}
\end{figure}
Fig.~\ref{fig:Lorenz_Att} shows that the $x$-$y$ plot for the ODE simulated (blue) and ESN predicted (orange) series, respectively, have attractors that are very similar in nature.

We note that the usual time-delay embedding methods, based on Takens' embedding theorem, can reconstruct a dynamical system up to a homeomorphism or diffeomorphism, i.e., the reconstructed coordinates are functions of the original dynamical system. In contrast, the ESN predicts coordinates that are identical to the original coordinates, specifically due to the fact that the training is done using the original coordinates.

In summary, RC can predict the short-term time series, but it fails to predict in the long run, shown in Fig.~\ref{fig:chua_test_pred} and Fig.~\ref{fig:Lorenz_Shrt_Pred}, as is expected for a chaotic system. However, RC can capture the long-term dynamics for both cases, which can be seen by a visual inspection of the simulated and predicted attractors in Fig.~\ref{fig:chua_attr} and Fig.~\ref{fig:Lorenz_Att}. 

\subsubsection{Statistical characteristics of Chua and Lorenz models}\label{ssec-stat-char-ode}
The results of the previous section suggest that the network is able to capture the statistical characteristics of the dynamics of the system. In order to quantify this, we use the following metrics: maximal Lyapunov exponent, sample entropy, 0-1 test for chaos, and the KDE plots.

\begin{table}[t!]
\caption{Comparison of maximal Lyapunov exponent (MLE), sample entropy (SE), and the asymptotic growth rate $K_c$ of the 0-1 test of the ESN predicted dynamics with those of the ODE simulations (top two rows) or the experimental Chau circuit (bottom row), for the various types of ESN (three panels)}
    \label{tab:table1}
    \centering

    Partial state input, full state output (Sec.~\ref{ssec-stat-char-ode},~\ref{subsec:experiment_chua's})

    \begin{adjustbox}{width=8cm}
    
    \begin{tabular}{|c|c|c|c|c|c|c|}
    \hline
    \multirow{2}{4em}{System}& \multicolumn{2}{c|}{MLE} & \multicolumn{2}{c|}{Sample Entropy} & \multicolumn{2}{c|}{$K_c$ for 0-1 test} \\ 
    \cline{2-7}   
      & Exp/Sim & Pred & Exp/Sim & Pred & Exp/Sim & Pred\\
     \hline\hline
     Lorenz 63 & 0.974 & 0.950 & 0.093 & 0.092 & 1.012 & 0.980\\
     \hline
     Chua ODE  & 0.105 & 0.147  & 0.082 & 0.082 & 0.758 & 0.803\\
     \hline
     Chua Experimental & 0.117 & 0.159 & 0.075 & 0.072 & 0.897 & 0.908\\
     \hline
    \end{tabular}
    \end{adjustbox}

    $\ $\\
    Full state input, full state output (Sec.~\ref{Sec:ESN_PH})
    $\ $\\

    \begin{adjustbox}{width=8cm}

    \begin{tabular}{|c|c|c|c|c|c|c|}
    \hline
    \multirow{2}{4em}{System}& \multicolumn{2}{c|}{MLE} & \multicolumn{2}{c|}{Sample Entropy} & \multicolumn{2}{c|}{$K_c$ for 0-1 test} \\ 
    \cline{2-7}   
      & Exp/Sim & Pred & Exp/Sim & Pred & Exp/Sim & Pred\\
     \hline\hline
     Lorenz 63 & 0.974 & 0.961 & 0.096 & 0.093 & 1.00 & 0.965\\
     \hline
     Chua ODE  & 0.105 & 0.093  & 0.082 & 0.082 & 0.754 & 0.773\\
     \hline
     Chua Experimental & 0.117 & 0.143 & 0.074 & 0.072 & 0.876 & 0.811 \\
     \hline
    \end{tabular}
    \end{adjustbox}

    $\ $\\
    Partial state input, partial state output (Sec.~\ref{subsec:partial_observation})
    $\ $\\

    \begin{adjustbox}{width=8cm}    
    \begin{tabular}{|c|c|c|c|c|c|c|}
    \hline
    \multirow{2}{4em}{System}& \multicolumn{2}{c|}{MLE} & \multicolumn{2}{c|}{Sample Entropy} & \multicolumn{2}{c|}{$K_c$ for 0-1 test} \\ 
    \cline{2-7}   
      & Exp/Sim & Pred & Exp/Sim & Pred & Exp/Sim & Pred\\
     \hline\hline
     Lorenz 63 & 0.974 & 0.956 & 0.093 & 0.092 & 0.991 & 0.999\\
     \hline
     Chua ODE  & 0.105 & 0.117  & 0.082 & 0.083 & 0.751 & 0.725\\
     \hline
     Chua Experimental & 0.117 & 0.1474 & 0.075 & 0.072 & 0.852 & 0.865\\
     \hline
    \end{tabular}
    \end{adjustbox}

\end{table}
Table.~\ref{tab:table1} (top panel) shows the maximal Lyapunov exponent (MLE), sample entropy (SE), and the asymptotic growth rate $K_c$ of the 0-1 test (see Ref.~\onlinecite[Section 3]{gottwald2009implementation}) for the simulated and predicted series for Lorenz 63 and Chua's oscillator.
We note that all these statistical properties of the simulated and predicted series are nearly equal.
\begin{figure*}[t!]
\centering
\includegraphics[angle=0,width=17.5cm]{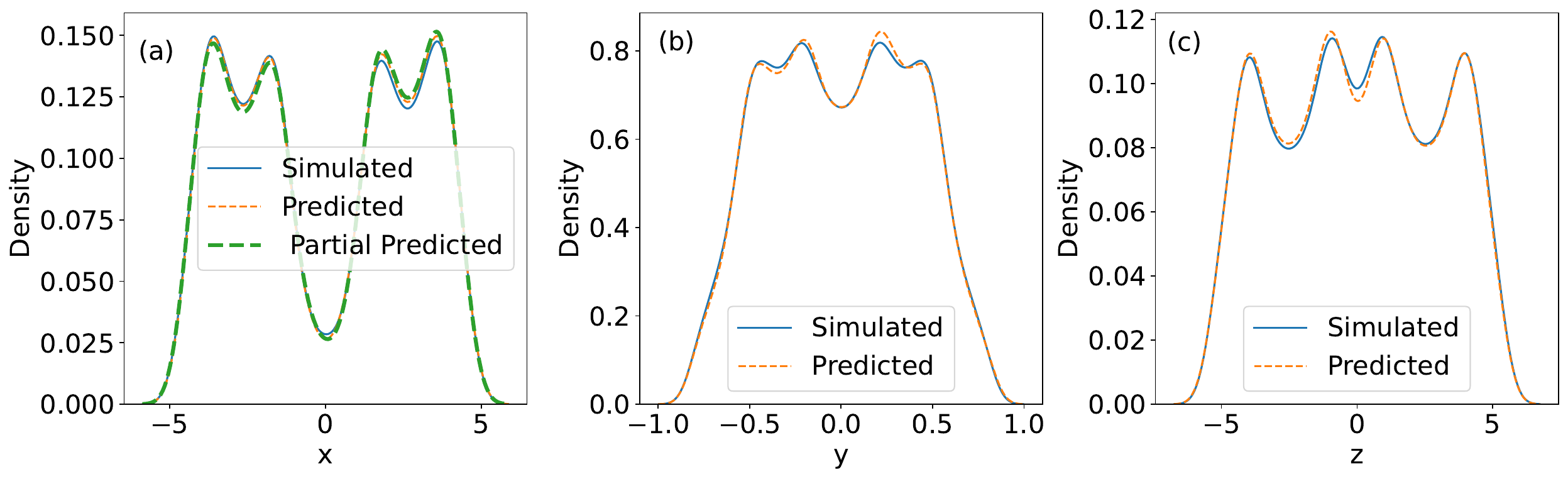}
\caption{\label{fig:chhu_pdf} Kernel density estimated (KDE) distributions for ODE simulated (solid) and ESN predicted data, for ESN with three-dimensional output (thin dashed) or ESN with one-dimensional output (thick dashed, left panel), for Chua ODE.}
\end{figure*}
Fig.~\ref{fig:chhu_pdf} shows the KDE plots of the simulated (solid) and ESN predicted (thin dashed) time series of Chua's oscillator for each of the three components of the system. 
The figures show that the statistical distribution of the simulated and predicted time series are similar in nature. A qualitatively similar result is also seen for Lorenz 63 system (figure for KDE plots not shown).
Thus, from these comparisons of various statistical measures, we see that the ESN can capture the long-term dynamics of the system accurately. 

\subsubsection{Study of MSE with time}\label{ssec-mse-ph}
The MSE is a commonly used metric to compare the performance of neural networks in time series prediction tasks. The MSE curve denotes the evolution of MSE with time. We have already defined the MSE in Eq.~\eqref{eq:mse-def}.

\begin{figure}[t!]
\centering
\includegraphics[angle=0,width=8.5cm]{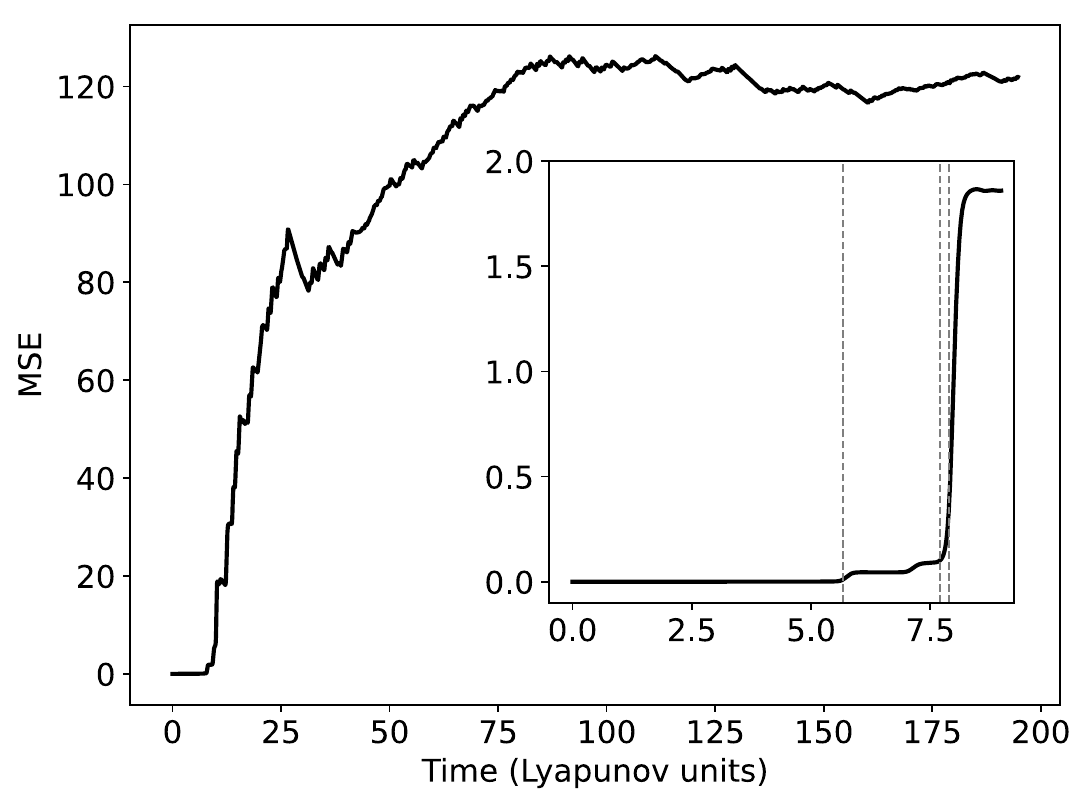}
\caption{\label{fig:MSE} Variation of MSE with time for Lorenz 63 model, comparing full state output ESN predictions with the ODE simulations. The dotted lines (from left to right) represent the prediction horizon for $r=$ 0.01, 0.1, and 0.3, respectively, as defined in Eq.~\eqref{eq:ph-r}.}
\end{figure}
Fig.~\ref{fig:MSE} shows the MSE curve over 200 Lyapunov times for Lorenz 63 model, for a typical trajectory. The MSE is close to 0 for a short period and increases rapidly thereafter, indicating the chaotic nature of the system, and eventually saturates. \textcolor{blue}{We observe that the level of saturation is indeed close to twice the variance of the attractor. A more precise explanation is provided in the Appendix~\ref{appendix:MSE_Saturation}.}
The zoomed-in plot shows the MSE curve for the initial period. The dotted lines (from left to right) represent the prediction horizon for $r=$ 0.01, 0.1, and 0.3, respectively, as defined in Eq.~\eqref{eq:ph-r}. The two series diverge quite quickly once the MSE crosses 0.3, implying that the short-term predictions are accurate only until that point. Despite the rapid increase, the MSE still remains bounded because both the test data and the predictions made by the ESN are bounded.

\begin{figure}[t!]
\centering
\includegraphics[angle=0,width=8cm]{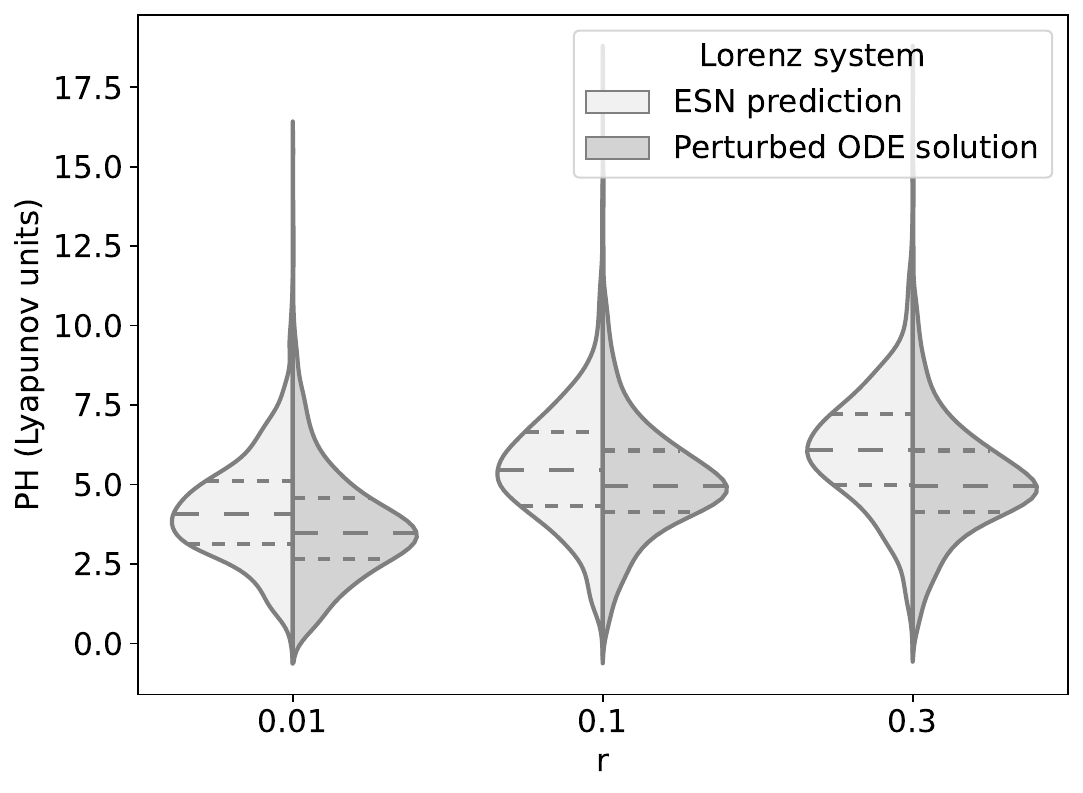}
\caption{\label{fig:ODE_ESN_comparison} Violin plots on the left show distributions of the prediction horizon $PH(0.01)$, comparing full state output ESN predictions with the ODE simulations. The right side shows distributions of the prediction horizon $PH(0.01)$, but for nearby trajectories with initial distance $\delta(0)=2.22\times 10^{-3}$. Both are for Lorenz 63 and use Lyapunov units.}
\end{figure}
\begin{table}[t!]
 \caption{Median prediction horizon $P(r)$, defined in Eq.~\eqref{median-ph}, in Lyapunov units, with different values of r, for the three systems: Lorenz 63, Chua ODE simulated, and Chua experimental system}
    \label{tab:TABLE2}
    \centering
    \begin{tabular}{|c|c|c|c|}
    \hline
     System & $P(0.01)$ & $P(0.1)$ & $P(0.3)$ \\
     \hline\hline
     Lorenz 63 & 4.1054 & 5.425 & 6.0875 \\
     \hline
     Chua ODE & 1.87 & 5.11 & 5.82\\
     \hline
     Chua Experimental & 2.173 & 3.067 & 3.48\\
     \hline
    \end{tabular}
\end{table}
The qualitative behaviour of the MSE as a function of time is very similar for different trajectories. But the time at which the MSE crosses a threshold $r$ varies quite a lot. We will discuss the details of this variation next.

\subsubsection{ESN predictability compared to uncertainty in initial condition}\label{sec:uncertainty_initcond}
Fig.~\ref{fig:ODE_ESN_comparison} (left side) shows the distributions of $PH(r)$ for three values of the threshold $r = 0.01, 0.1, 0.3$ for the Lorenz 63 system, using a sample size of 1000. We note that the distributions shown in these violin-plots demonstrate the variation of the prediction horizon with respect to the initial condition on the attractor, \textcolor{blue}{in contrast with the variability due to variations in the prediction model, as has been discussed in previous works:
Ref.~\onlinecite{vlachas2020backpropagation} shows the variations of `valid prediction time' across different hyperparameters of the ESN (figure~4 in Ref.~\onlinecite{vlachas2020backpropagation}), while Ref.~\onlinecite{GOTTWALD2021132911} discuss the variation of `forecast time' over different realizations of their model (random feature map) each trained on its own training set (figure~1 in Ref.~\onlinecite{GOTTWALD2021132911}). 
We note that the variability due to varying initial conditions is significantly larger than that due to varying models - the spread of the distributions in Fig.~\ref{fig:ODE_ESN_comparison} is more than that in the figures in Ref.~\onlinecite{vlachas2020backpropagation, GOTTWALD2021132911} quoted above.}

We see that with varying initial conditions, the prediction horizon can range from values very close to 0 (trajectories that are very difficult to predict using the ESN) to as large as 15 or more Lyapunov times (for trajectories with very high predictability). One way to understand this distribution is to compare with the predictability of the original system itself, as we discuss below.

As is well known, for chaotic systems, a small initial uncertainty $\delta(0)$ shows exponential growth asymptotically in time, with the average growth rate determined by the Lyapunov exponent $\lambda$. Heuristically, the expression for $\delta(t)$ is, on an average, given by the equation
\begin{equation*}
    \delta(t)\approx \delta(0)e^{\lambda t} \,.
\end{equation*}
It is also well known that the actual (not average) behaviour of $\delta(t)$ shows substantial variation based on the initial condition in the phase space. The value of time at which $\delta(t)^2$ crosses a predetermined threshold thus depends a lot on the initial condition.
Fig.~\ref{fig:ODE_ESN_comparison} (right side, shaded violin plots) shows the distributions of these times for a sample of 1000 trajectories, with initial distance $\delta(0)=2.22\times 10^{-3}$. With varying $\delta(0)$, the median of the shaded distributions on the left in Fig.~\ref{fig:ODE_ESN_comparison} varies but the shapes remain qualitatively similar. \textcolor{blue}{A detailed description of the variation of the PH distribution as well as an explanation for the choice of $\delta(0)$ is now provided in Appendix~\ref{appendix:ph_pert}}. As expected, we again see a wide distribution where some neighbouring trajectories do not diverge beyond the threshold $r = 0.01$ for as long as 15 or more Lyapunov time units, whereas some do so at times very close to 0.

We note that the shape and importantly the support of both these distributions -- one for the ESN prediction horizon and other for the divergence times as explained above -- are very similar to each other. Indeed, the initial distance $\delta(0)=2.22\times 10^{-3}$ was chosen so that the two distributions (on the left and right) ``match'' closely. The interpretation is that the ESN predictions are equivalent to those of the true system (ODE solver) with an uncertainty of the order $\delta(0)\approx10^{-3}$ in the initial conditions. 

\subsubsection{Effect of Noisy training data}\label{ssec-noise}
\begin{figure}[t!]
\centering
\includegraphics[angle=0,width=8.5 cm]{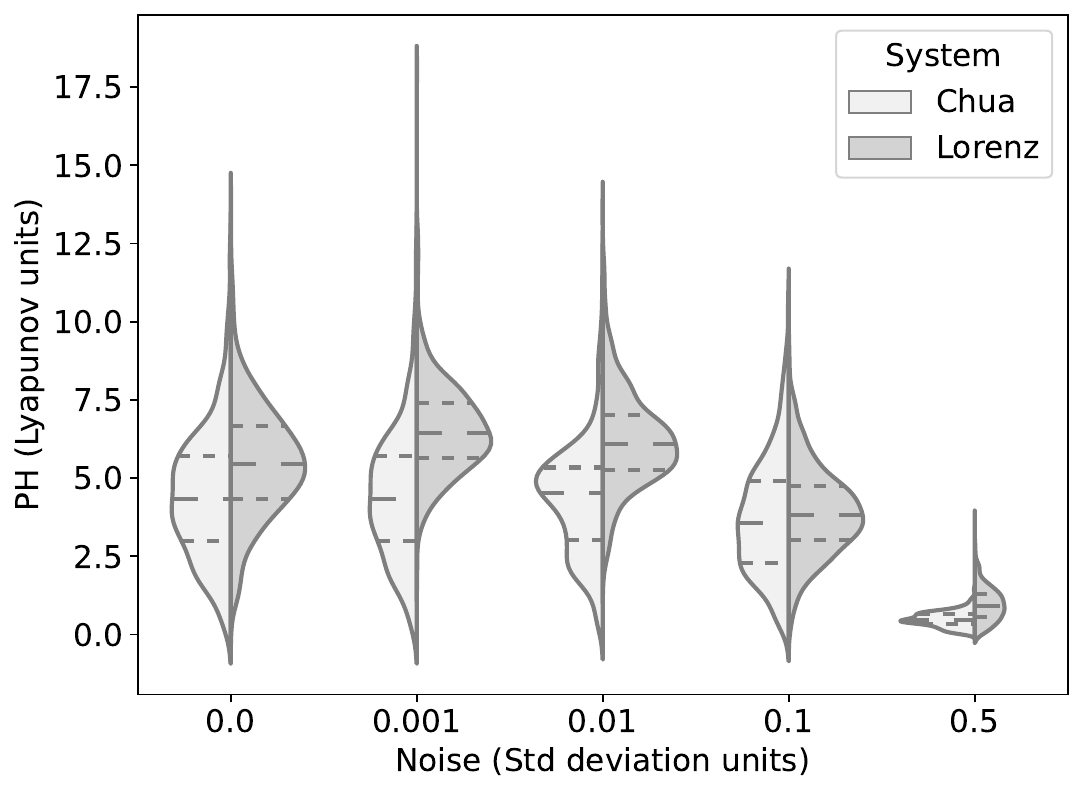}
\caption{\label{fig:PH_noise} Distributions of prediction horizon with different noise levels for the Chua oscillator (left side) and Lorenz 63 system (right side).}
\end{figure}
Finally, we test to see if the ESN is robust to noisy datasets, i.e., whether the ESN can pick up the essential dynamics of the system despite the noise. Fig.~\ref{fig:PH_noise} shows the variation of the prediction horizon for the cases when the datasets have different levels of Gaussian noise added to the training data, i.e., simulated trajectories of the Chua's and Lorenz 63 systems with noise added. Note that we are not simulating a stochastic differential equation but just adding noise to the deterministic trajectory, similar in spirit to the use of noisy observations of a deterministic system in many application areas such as the earth sciences.

The left and right sides of each plot indicate the Chua's and Lorenz 63 systems, respectively. The median prediction horizon along with the 25th and 75th percentiles are shown by horizontal dashed lines in each plot in Fig.~\ref{fig:PH_noise}. We see the median remains nearly constant for small amounts of noise, but when the noise amplitude is large, the ESN finds it difficult to learn the dynamics. Thus, the study reveals that ESN can capture the essential dynamic with a certain amount of noise, but it fails with higher noise levels.

\subsubsection{\label{subsec:experiment_chua's} ESN trained using experimental data for Chua's oscillator}
\begin{figure}[t!]
\centering
\includegraphics[angle=0,width=8.5cm]{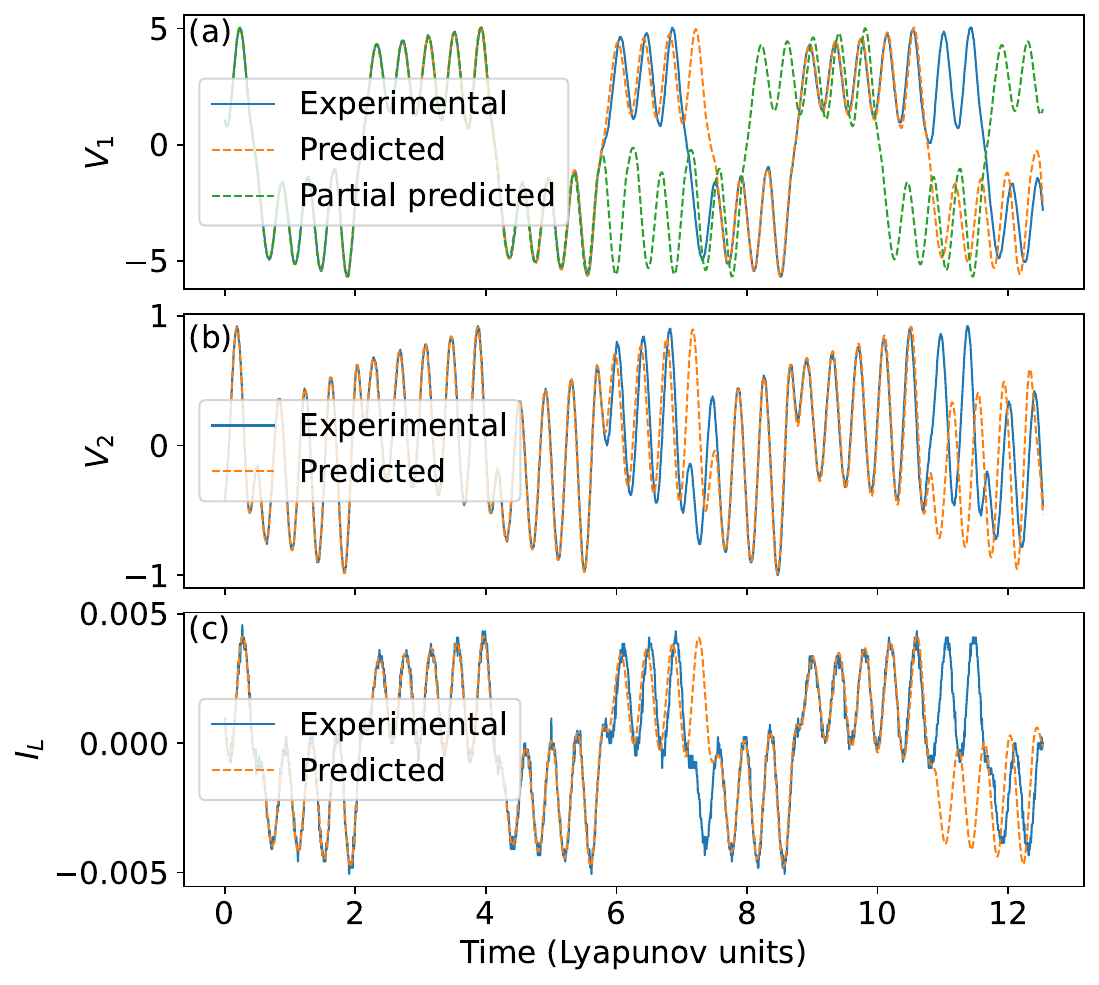}
\caption{\label{fig:Chua_exp_pred_test} (a)-(c) represent the experimental and predicted time series for the $V_1$, $V_2$ and $I_L$ variables of the experimental Chua's circuit. The thick line in plot (a) represents the partially predicted series of Chua's experimental data.}
\end{figure}
We now discuss the performance of ESN in emulating the dynamics of the Chua circuit when trained and tested with experimental data. We recall that the ESN input is one dimensional (the voltage $V_1$) while the output is three dimensional -- $V_1, V_2, I_L$ dimensional variables (no pun intended).

As with the previous section, Fig.~\ref{fig:Chua_exp_pred_test} shows a sample trajectory of the experimental (solid) and ESN-predicted (thin dashed) time series of $V_1$, $V_2$, and $I_L$, respectively. Note that in this case, we use the dimensional variables instead of the dimensionless variables. Further, these are not scaled in any way and the numerical ranges of the voltage and current variables are vastly different. The median prediction horizon $P(0.01)$ is 2.173 Lyapunov times, very similar to the case of ESN trained using ODE simulated data. 

Fig.~\ref{fig:chua_attr}(b) shows the double scroll attractor ($V_1$-$V_2$) for the experimental series. Fig.~\ref{fig:chua_attr}(c) blue and orange colors show the attractor using ESN trained with experimental data and ESN trained with ODE simulation data, respectively. The dissimilarity between the experimental and ODE-simulated attractors is attributed to the noisy experimental data. The attractor from the experimental data is not only noisy but also clearly shows the effect of the low resolution (compared to numerics) of the experimental measurements. But a striking feature is that the ESN predicted attractor shown in blue color in Fig.~\ref{fig:chua_attr}(c) is, in fact, able to capture the details that are not visible in the experimental dataset shown in Fig.~\ref{fig:chua_attr}(b).

\begin{figure*}[t!]
\centering
\includegraphics[angle=0,width=17cm]{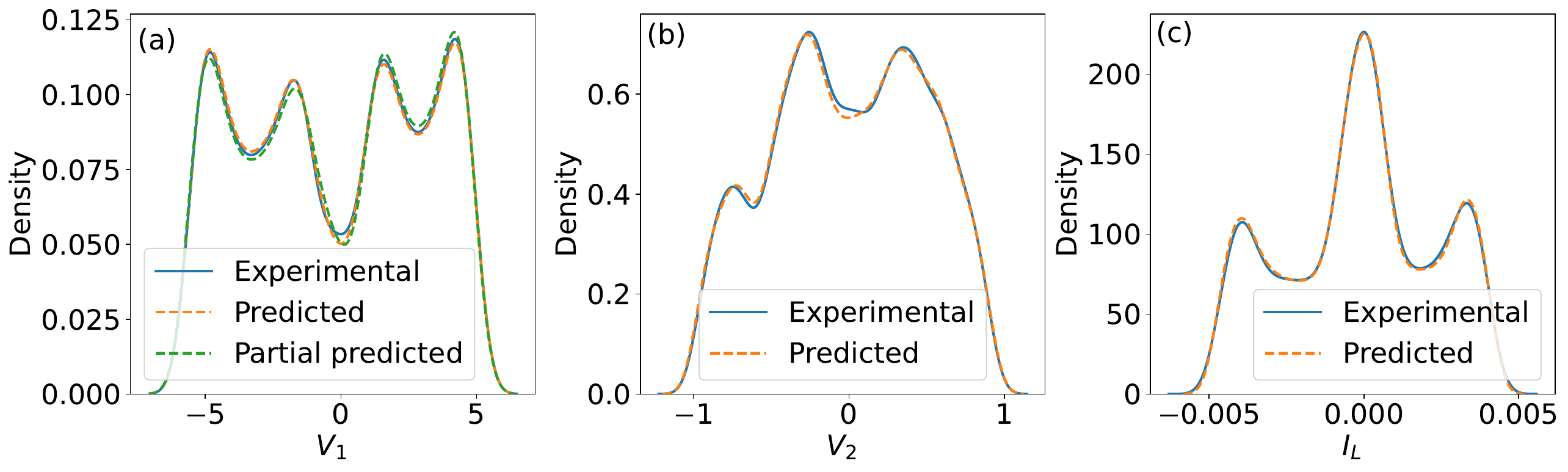}
\caption{\label{fig:Chua_exp_pdf} Same as Fig.~\ref{fig:chhu_pdf} but with experimental data instead of ODE simulated data.}
\end{figure*}
Table~\ref{tab:table1}, bottom row of the top panel, lists the comparison of the various statistical measures, namely, the maximal Lyapunov exponent, sample entropy, and the asymptotic growth rate for the 0-1 test of chaos, while Fig.~\ref{fig:Chua_exp_pdf} shows the KDE plots for the experimental (solid) and ESN predicted (dashed) time series. Very similar to the previous case of training using simulated data, we see that the ESN trained using experimental data is also able to capture the statistical properties of Chua circuit dynamics.

\begin{figure*}[t!]
\centering
\includegraphics[angle=0,width=17.5cm]{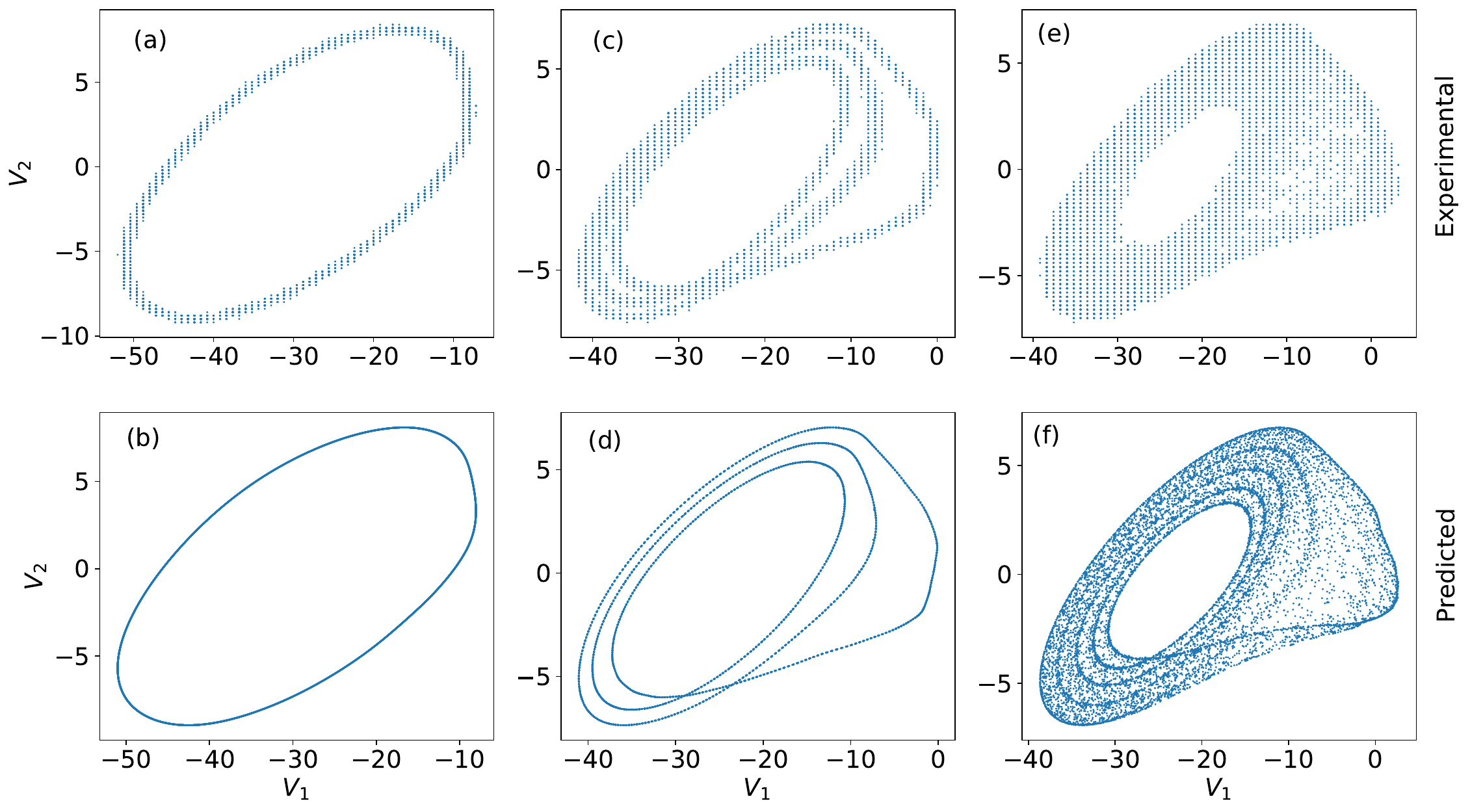}
\caption{\label{fig:Chua_exp_dyn} Chua's circuit with different dynamics: top row shows experimental data, and the bottom row shows ESN predictions for period one, period three, and one scroll chaotic dynamics (from left to right, respectively).}

\end{figure*}
Besides studying the double scroll attractor, we have also studied different dynamics - periods one, three, and single scroll of Chua's oscillator. Fig.~\ref{fig:Chua_exp_dyn} (a), (c), (e), and (b), (d), (f) represent, respectively, the experimental and ESN predicted dynamics for period one, period three, and a single scroll. We again see the remarkable result that even though the experimental measurements are, as expected, noisy and of a low resolution, the ESN trained using these noisy measurements is able to capture and predict the Chua circuit dynamics with high fidelity. A thorough theoretical investigation of this property of ESN will be a fruitful avenue for further research.

\subsection{\label{Sec:ESN_PH} Comparison of partial state input ESN with full state input ESN}

\begin{figure}[t!]
\centering
\includegraphics[angle=0,width=8cm]{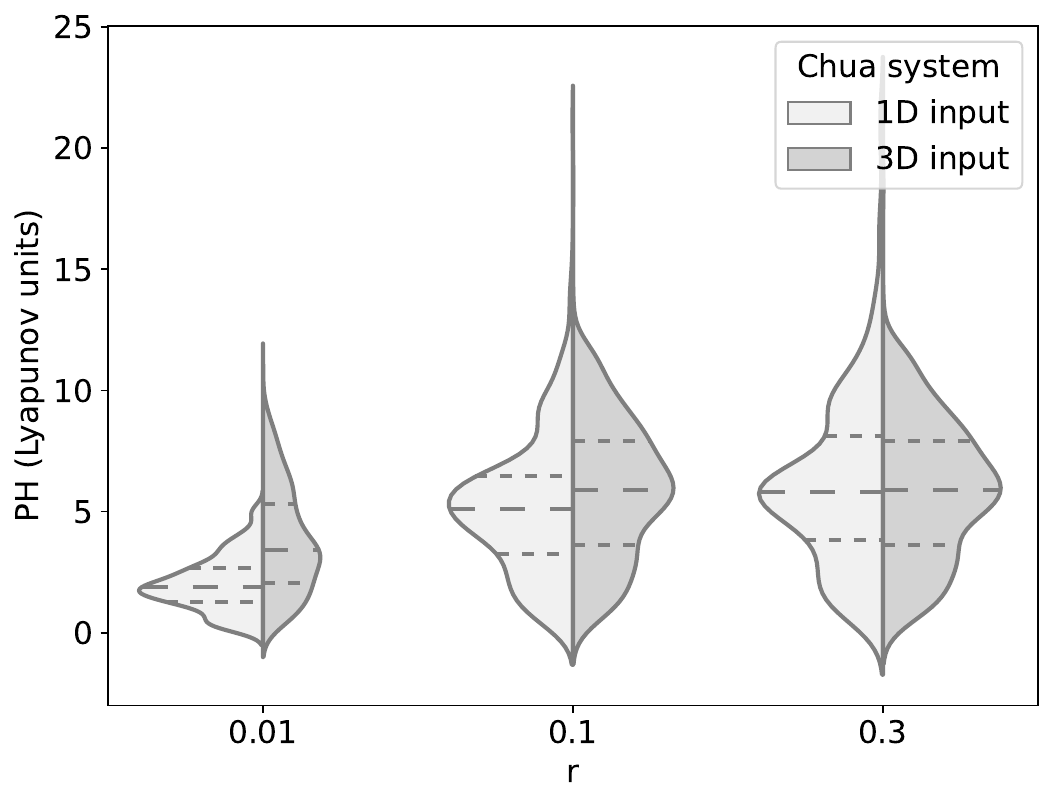}
\caption{\label{fig:chua_violin_PH} Distributions of the prediction horizon $\text{PH}_j(r)$ for Chua ODE for two types of ESNs - one with one-dimensional partial state input (left side) and the other with three-dimensional full state input, for three different values of the threshold $r$.}
\end{figure}
\begin{figure}[t!]
\centering
\includegraphics[angle=0,width=8cm]{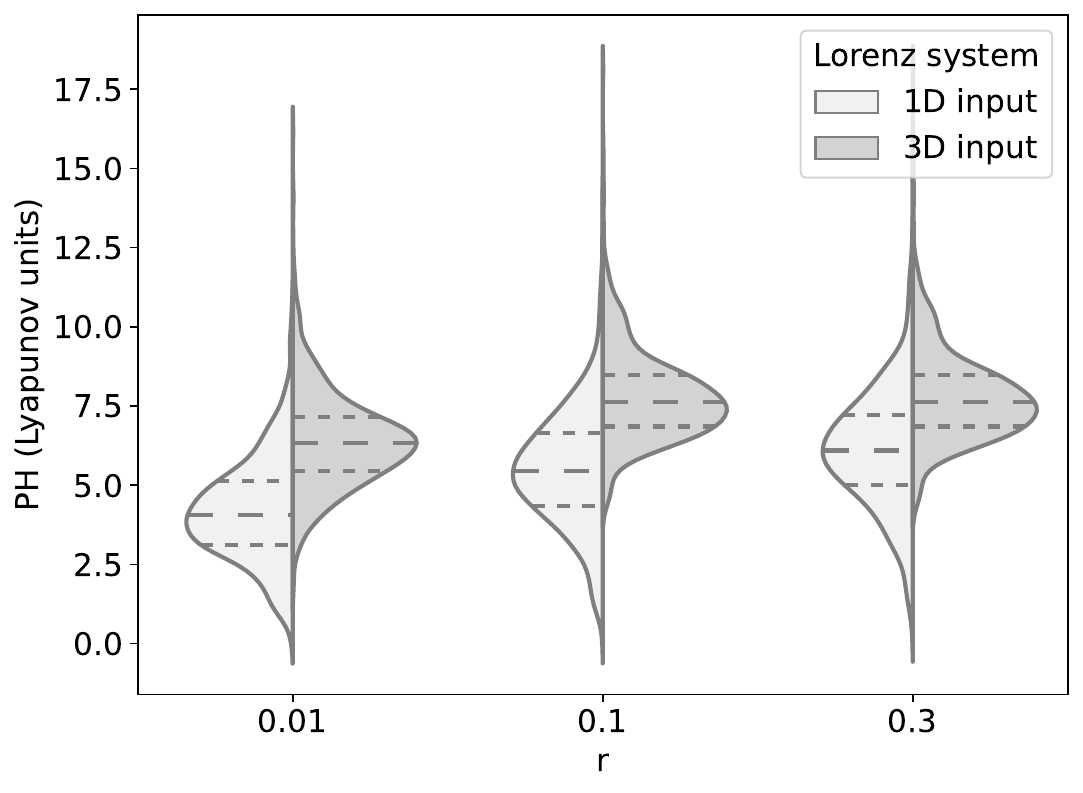}
\caption{\label{fig:L63_violin_PH} Same as Fig.~\ref{fig:chua_violin_PH} but for Lorenz 63 model.}
\end{figure}
As discussed extensively in the previous sections, the ESN trained with one-dimensional partial state input and three-dimensional full state output is indeed able to model the dynamics of the full system quite well. Thus it is expected that the ESN can accomplish the same task using the full state input, as we now discuss. Thus, in this case, $\mathbf u(t) = (x(t), y(t), z(t)) \in \mathbb R^3$ and $\mathbf y(t) = (x(t), y(t), z(t)) \in \mathbb R^3$. ESNs with such a setup have been discussed in detail in previous studies.~\cite{pathak2017using,haluszczynski2019good}. Hence this section is very brief and only discusses the comparison of full-state vs.~partial-state input ESN using the distribution of the prediction horizon which has not been considered in previous studies.

Since more input data is being provided to the ESN when trained with full-state input, it may be expected that its predictions will be more accurate than the partial-state input case. Fig.~\ref{fig:chua_violin_PH} and Fig.~\ref{fig:L63_violin_PH} show a comparison between the prediction horizon distribution for partial state input ESN (left side) and full state input ESN (right side). We see that the predictability of the full state input ESN is superior to that of the partial state input ESN for the case of the Lorenz 63 system (Fig.~\ref{fig:L63_violin_PH}), but there is very little improvement in the case of the Chua oscillator (Fig.~\ref{fig:chua_violin_PH}). It will be an interesting avenue of future research to investigate which dynamical characteristics of the Chua oscillator and Lorenz 63 model lead to such a difference and to provide a precise characterization of the kind of systems for which the partial and full state input ESN perform equally well as compared to the systems for which they do not perform equally well.

\subsection{\label{subsec:partial_observation} Comparison of partial state output ESN with full state output ESN}

The other natural variation of the ESN will be to consider the case when both the input and output are partial states and not full states. Note that all the results in both the previous Sec.~\ref{subsec:Cha_ckt} and Sec.~\ref{Sec:ESN_PH} are for ESN with three-dimensional full state output, i.e., $\mathbf y(t) = (x(t), y(t), z(t)) \in \mathbb R^3$, while in this section, we consider the case of one-dimensional partial state output: $\mathbf y(t) = x(t) \in \mathbb R$.

We report some of these comparisons below.
\begin{enumerate}
    \item Panels (a) of Fig.~\ref{fig:chua_test_pred}, Fig.~\ref{fig:Lorenz_Shrt_Pred}, and Fig.~\ref{fig:Chua_exp_pred_test} show the time series (thick dashed) of ESN predicted variable $x$ (or $V_1$) of the Chua ODE, Lorenz 63 model, and the Chua experimental cases, respectively. Of course, in this case, we cannot compare the model attractor (without using some additional technique, such as time delay embedding, which is not the focus of this paper).
    \item Similarly, panels (a) of Fig.~\ref{fig:chhu_pdf} and Fig.~\ref{fig:Chua_exp_pdf} show the KDE plot (thick dashed) of the variable $x$ of the ESN simulation of Chua ODE, and of $V_1$ of Chua's experimental system, respectively.
    \item Further, the bottom panel of Table~\ref{tab:table1} shows a comparison of other statistical quantities, namely, the maximal Lyapunov exponent, the sample entropy, and the asymptotic growth rate for the 0-1 test.
    \item It is also found that the prediction horizon is 3.125 Lyapunov times for Lorenz 63 partial variable $x$, 2.29 Lyapunov times for Chua's partial variable $x$, and 1.44 Lyapunov times for Chua's experimental variable $V_1$. (Plots of distributions of prediction horizon are not shown here.)
\end{enumerate}
We see that in all these qualitative and quantitative metrics, the performance of the partial state output ESN is very similar to the case of full state output ESN.

\section{\label{sec:con} Conclusion}
In this study, we propose an echo state network (ESN) based approach for reconstructing the full state of a dynamical system from its partial observation. We demonstrate effectiveness of this framework with two examples: the Lorenz system and Chua's oscillator, including the use of experimental data for the latter system. We also provide a heuristic justification for our paradigm. A major contribution is a thorough investigation of the variability of the prediction horizon of a dynamical system across several initial conditions. 

We have demonstrated that ESN can predict the short-term time series up to a few Lyapunov times but fails eventually. However, there is large variability in the prediction horizon for different initial conditions. The distribution of prediction horizon values over many initial conditions is studied in Sec.~\ref{sec:uncertainty_initcond}, which we believe is a better way of quantifying the short-term predictability of the ESN. The similarity of this distribution with that of the time for divergence of nearby trajectories, as shown in Fig.~\ref{fig:ODE_ESN_comparison}, shows that this variability is an inherent characteristic of chaotic systems and not just a consequence of or a property of the use of ESN.

A comparison of the predicted attractor with the simulated attractor seems to suggest that the ESN successfully replicates the system's long-term statistics. Several metrics, namely, the maximal Lyapunov exponent, sample entropy, the asymptotic growth rate of the mean square displacement used in the 0-1 test of chaos, as well as the kernel density estimates of marginal distributions of the dynamical variables described in Sec.~\ref{sec:s_test} have been used to quantify the results. The estimated values of these metrics for the ESN dynamics match closely with those obtained from the simulated or experimental data, providing strong evidence that the ESN can accurately capture the long-term dynamics, even when trained on noisy data. 

In Sec.~\ref{Sec:ESN_PH}, we compare our framework with the more commonly studied full-state input, full-state output schemes and we observe that the prediction horizon distribution and statistical measures are comparable for the two schemes. As we mentioned earlier, the performance of the ESN depends on the choice of the spectral radius $\rho$, reservoir size $N$, $\mathbf{W}^{\textup{in}}$, $\mathbf{W}$, and are chosen through a trial and error process. A detailed mathematical study of the effect of these choices on the performance of the ESN and ways to optimise these choices would be an interesting direction of future research.

Finally, in Sec.~\ref{ssec-noise}, we observe that the ESN is able to capture the dynamics of the system very well with low noise levels, but as expected, with increasing magnitude of the noise, the prediction horizon reduces significantly. We also use noisy experimental data to train the ESN. In this case, the ESN is indeed able to capture the experimental attractor's dynamical characteristics quite well. This aspect of the ESN makes it an invaluable tool for application in the prediction of real-world datasets, at least in low-dimensional settings that we have studied. Applications of our framework to study high-dimensional dynamics as well as developing a theoretical understanding of the ability of ESN to `filter' the noise will be fruitful directions of future research.

\section*{Acknowledgment}
 We would like to acknowledge the UG-Physics and Atomic Physics and Quantum Optics Lab, IISER Pune, for allowing us to conduct the experiments and would like to thank Korak Biswas for helping with the experimental work. 

\bibliography{RC_manus.bib}   

\color{blue}{
\appendix

\section{}

\subsection{Effect of reservoir size on prediction horizon}\label{appendix:size_ESN}
\begin{figure}[t!]
\centering
\includegraphics[angle=0,width=8.5cm]{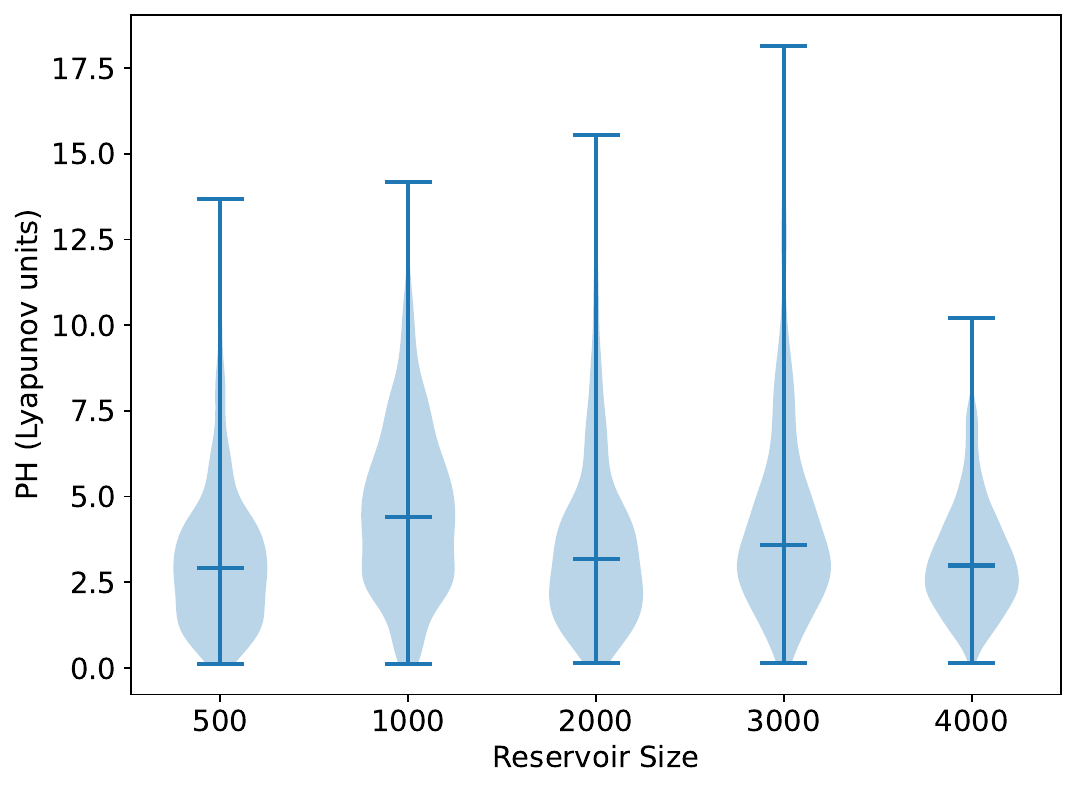}
\caption{\label{fig:Chua_PH_RSize} \textcolor{blue}{Distribution of the prediction horizon with reservoir size.}}
\end{figure}
The effect of varying the reservoir size is depicted in Fig.~\ref{fig:Chua_PH_RSize}, which shows the distributions of prediction horizons in the case of partial observation of Chua’s oscillator. We note that these distributions are not significantly different - the differences in the tails and quartiles are unlikely to have
statistical significance. 

\subsection{Variation of PH with perturbation for ODE solver}\label{appendix:ph_pert}
\begin{figure}[t!]
\centering
\includegraphics[angle=0,width=9.2cm]{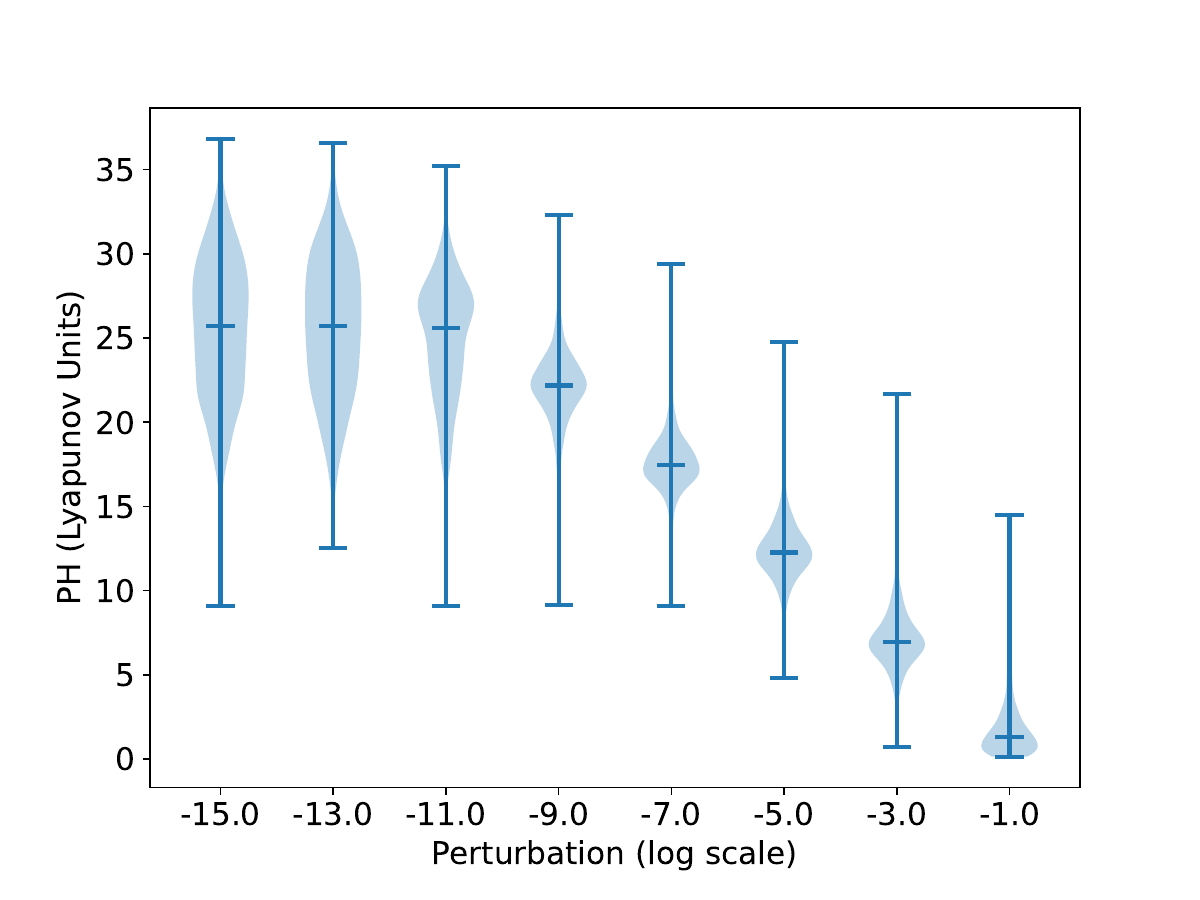}
\caption{\label{fig:PH_pert_ODE_solver} \textcolor{blue}{ Distribution of the prediction horizon with perturbation of initial condition for the ODE solver}}
\end{figure} 
We show the variation of the $PH$ as a function of the perturbation in the initial condition for the ODE solver in Fig.~\ref{fig:PH_pert_ODE_solver}. We notice a linear trend in the median of the PH distribution which we explain using the following argument. For chaotic systems, we expect an initial error $\delta(0)$ to grow as per,
$$\delta(t)=\delta(0)e^{\lambda t}.$$
Let $\hat{t} = \lambda t$ denote the time (in Lyapunov units) at which the MSE exceeds the threshold 0.1 and denote $K = \log\delta(\hat{t})$, then we have \begin{align*}
    K&=\log\delta(0)+{\hat{t}},\\
    \hat{t}&=\left(K-\log \delta(0)\right),\\
    &=-\log(10) \ \log_{10}(\delta(0)) + K.
\end{align*}
We expect the slope of $PH$ vs $\log_{10}\delta(0)$ to have slope $m=-\log 10\approx-2.302$. The slope of the graph in Fig.~\ref{fig:PH_pert_ODE_solver}, calculated by ignoring the values at -13 and -15, is $m=-2.459$ which is fairly close to the expected slope of -2.302.

Fig.~\ref{fig:PH_pert_ODE_solver} also provides a clear explanation for the choice of the initial perturbation $\delta(0) = 2.22 \times 10^{-3}$ used for the distributions in Fig.~\ref{fig:ODE_ESN_comparison} - the main aim was to match the PH distribution for the ESN predictions.

\subsection{Time complexity}\label{appendix:TC}
We compute the time complexity of training the ESN in terms of size of the training dataset $T$, the size of the reservoir $N$, dimension of the dynamical system $n$, and the dimension of the ESN input $m$, assuming $T \gg N \gg m, n$. There are two main steps during training. (i) Calculation of $\mathbf X$: The calculation of $x(t)$ takes $\mathcal{O}(N(N+m))$ and hence calculating $\mathbf X$ takes  $\mathcal{O}(TN(N+m))$. (ii) Calculation of $\mathbf{W}^\text{out}$ in Eq.~\ref{eq:wout}: The matrix inversion and multiplication takes $\mathcal{O}(T(N+m)^2)$. Thus, the total time complexity of the training is $\mathcal{O}(TN^2)$.

\subsection{MSE saturation level}\label{appendix:MSE_Saturation}
\begin{table}[h]
    \centering
    {\color{blue}\begin{tabular}{|c|c|c|}
    \hline
     & ESN predictions   &  Actual trajectory\\
     \hline
    $\mu_x$ & -0.166 & -0.08\\
    $\mu_y$ & -0.166 & -0.086\\
    $\mu_z$ & 23.4  & 23.5\\
    \hline
    
    \end{tabular}}
    \caption{\textcolor{blue}{Mean of the ESN and true trajectories (used to calculate the MSE reported in Fig~\ref{fig:MSE})}}
    \label{tab:avgtraj}
\end{table}
We provide here an explanation of the saturation level of the MSE seen in Fig.~\ref{fig:MSE}. We denote the $x$ component of the ESN predicted trajectory at time $i$ as $x_{i,1}$ and the actual trajectory at time $i$ as $x_{i,2}$. We notice in Table~\ref{tab:avgtraj} that the sample mean of the ESN predicted trajectory and actual trajectory are nearly equal, which we denote as the mean of the attractor by $\bar{x} \approx -0.12$. 
We also compute the covariances for these two samples, $\{ x_{i,1}\}_{i=1}^N$ and $\{ x_{i,2}\}_{i=1}^N$, which are as follows:
\begin{align}
    \Sigma_x &= \left(\begin{array}{cc}
    62.4     &  10.4\\
    10.4 & 62.6\\
    \end{array}\right) \,,
    \,
    \Sigma_y = \left(\begin{array}{cc}
     81.7    &  13.0\\
    13.0 & 81.4\\
    \end{array}\right) \,, \label{sigmas}
    \, \\
    \Sigma_z &= \left(\begin{array}{cc}
     76.3     &  13.4\\
    13.4 & 75.0 
    \end{array}\right) \,. \notag
 \end{align}
We notice that (i) the covariances are indeed nonzero and (ii) the variances (diagonal elements) are nearly equal.

The MSE can now be computed as follows:
\begin{align}
    MSE_x&=\frac{1}{n}\sum_{i=1}^n (x_{i,1}-x_{i,2})^2 \notag \\
    &=\frac{1}{n}\sum_{i=1}^n ((x_{i,1}-\bar{x})-(x_{i,2}-\bar{x}))^2 \notag \\
    &={\sigma_{x_1}}^2+{\sigma_{x_2}}^2-2\text{cov}(x_{i,1},x_{i,2}) \label{mse-sigma} \\
    &=2\sigma_x^2 \quad \textrm{only if} \quad \text{cov}(x_{i,1},x_{i,2}) = 0 \notag
\end{align}
Thus using~\eqref{sigmas} in~\eqref{mse-sigma}, we get
\begin{align}
    MSE_x = 104 \,, \qquad
    MSE_y = 137 \,, \qquad
    MSE_z = 124 \,, \qquad
\end{align}
and the MSE which is the average of the above three is obtained as $MSE = 120$ which is close to the level of saturation seen in Fig.~\ref{fig:MSE}.}

\end{document}